\documentclass[twocolumn]{aastex62}

\usepackage{graphicx}
\usepackage{dcolumn}
\usepackage{bm}
\usepackage{color}
\usepackage{xcolor}
\usepackage{amsmath}
\usepackage{lineno}

\newcommand\lsim{\mathrel{\rlap{\lower4pt\hbox{\hskip1pt$\sim$}}
\raise1pt\hbox{$<$}}}
\newcommand\gsim{\mathrel{\rlap{\lower4pt\hbox{\hskip1pt$\sim$}}
\raise1pt\hbox{$>$}}}
\graphicspath{{./}{figures/}}
\revised{\today}
\submitjournal{ApJ}

\begin{document}

\title{From Wide Triples to UCXBs: Multimessenger Signatures of Dynamically-formed Black Hole-White Dwarf Systems in the LISA Band}




\author{Zeyuan Xuan}
\affiliation{ Department of Physics and Astronomy, UCLA, Los Angeles, CA 90095}
\affiliation{Mani L. Bhaumik Institute for Theoretical Physics, Department of Physics and Astronomy, UCLA, Los Angeles, CA 90095, USA}

\author{Cheyanne Shariat}
\affiliation{Department of Astronomy, California Institute of Technology, 1200 East California Boulevard, Pasadena, CA 91125, USA}

\author{Smadar Naoz}
\affiliation{ Department of Physics and Astronomy, UCLA, Los Angeles, CA 90095}
\affiliation{Mani L. Bhaumik Institute for Theoretical Physics, Department of Physics and Astronomy, UCLA, Los Angeles, CA 90095, USA}

\begin{abstract}
Ultracompact X-ray binaries (UCXBs) are a subclass of low-mass X-ray binaries characterized by tight orbits and degenerate donors, which pose significant challenges to our understanding of their formation. Recent discoveries of black hole (BH) candidates with main-sequence (MS) or red giant (RG) companions suggest that BH–white dwarf (BH-WD) binaries are common in the Galactic field.
Motivated by these observations and the fact that most massive stars are born in triples, we show that wide BH-WD systems can naturally form UCXBs through the eccentric Kozai-Lidov (EKL) mechanism. Notably, EKL-driven eccentricity excitations combined with gravitational wave (GW) emission and WD dynamical tides can effectively shrink and circularize the orbit, leading to mass-transferring BH-WD binaries. These systems represent promising multimessenger sources in both X-ray and GW observations. Specifically, we predict that the wide triple channel can produce {$\sim3-27$ ($\sim1-5$)} detectable UCXBs in the Milky Way (Andromeda galaxy), including $\sim1$ system observable by the mHz GW detection of LISA. If the final WD mass can reach sufficiently small values, this channel could contribute up to $\sim 10^3$ UCXBs in the Galaxy. Furthermore, the identification of tertiary companions in observed UCXBs would provide direct evidence for this formation pathway and yield unique insights into their dynamical origins.
\end{abstract}


\section{Introduction} \label{sec:intro}
Ultracompact X-ray binaries (UCXBs) are a subclass of low-mass X-ray binaries (LMXBs) with orbital periods shorter than $\sim 1$ hr \citep[][]{Savonije1986}. These systems host a neutron star or black hole accreting material from a compact, degenerate companion \citep[e.g., a white dwarf; see][]{Nelson1986,in_t_Zand_2007, Armas_Padilla_2023}. Due to the short orbital period and high mass-transfer rate, UCXBs are luminous X-ray emitters \citep[luminosity $L_X\sim10^{39}~{\rm erg~s^{-1}}$; e.g.,][]{Maccarone07} and strong candidates for mHz gravitational wave (GW) observation \citep[e.g.,][]{Qin_2023,Chen25}. Studying their observable properties can place robust constraints on compact-object formation and the physics of mass transfer. Furthermore, understanding their population demographics is especially important for joint X-ray and GW studies in a multi-messenger framework, while also enabling the development and validation of data analysis pipelines for future space-based GW detectors, such as LISA, TianQin, and Taiji \citep[e.g.,][]{amaro17,amaro+22,luo16,Ruan_2020}.


Recent observations have confirmed $\sim 20$ UCXBs in the Milky Way, with the majority hosting a neutron star accretor \citep[see, e.g.,][]{Armas_Padilla_2023}. Additionally, some systems exhibit evidence for a black hole accretor (BH-UCXB). For example, 47 Tuc X9, a bright X-ray source in the globular cluster (GC) 47 Tucanae with a 28-minute orbital period, is among the most promising  BH-UCXB candidates \citep{MillerJones2015, Bahramian2017, Church2017, Tudor2018}. More recently, a BH-UCXB candidate in the Andromeda galaxy has been reported, with an exceptionally short orbital period of only 7.7 minutes \citep[e.g.,][]{Zhang2024,yang2025}.

Despite these discoveries, the formation channels of UCXBs, {\it especially those with black hole accretors,} remain poorly understood. A common explanation of UCXB formation is the isolated evolution of main-sequence binaries \citep[e.g.,][]{Podsiadlowski2002, van_Haaften_2012}, in which two main-sequence stars undergo one or two common-envelope (CE) phases, and possibly a supernova, to form a compact-object binary in a tight orbit. However, BH progenitors (main sequence stars with mass $M\gsim20~{\rm M_\odot}$) can expand to $\sim 1000-3000~R_\odot$ during their stellar evolution \citep[e.g.,][]{Levesque2005,Romagnolo2023}. Consequently, any companion star within $\lesssim 10$~au will undergo mass transfer with the BH progenitor. If the secondary star is significantly less massive, as is the case for BH-UCXB progenitors, unstable mass transfer will be initiated, whereby the secondary spirals through the primary's envelope in a CE event. Given the extreme mass ratio at this stage ($q\lsim0.1$), the most probable outcome of CE is a merger, as the available orbital energy is often insufficient to completely eject the primary's envelope. {To overcome this difficulty, alternative energy sources for CE ejection have been proposed, see, e.g., \citet{Podsiadlowski2010,Ivanova2011,Ivanova2015}. These studies further suggest that additional contributions, such as thermonuclear or recombination energy, may assist envelope ejection, implying that the isolated binary channel, while highly challenging, cannot be entirely ruled out in the formation of BH-UCXBs.}


On the other hand, a natural resolution for BH-UCXB formation arises by considering dynamical formation channels. For example, studies have shown that globular clusters can efficiently produce BH-UCXBs through direct collisions, tidal captures, and exchange interactions \citep[e.g.,][]{Rasio_2000, Ivanov2005,Lombardi_Jr__2006,Ivanova_2010}. Moreover, since most ($\approx70\%$) BH progenitors are born in triple systems \citep[][]{Moe+17,Sana+12,Sana14,Offner23}, three-body dynamics are an inescapable consideration for the evolution of BHs in the galaxy. In particular, effectively all triple star systems in the field initially reside in hierarchical configurations
\citep[e.g.,][]{Tokovinin+08,Tokovinin22,Shariat25_10ktriples}, where a close inner binary is orbited by a distant tertiary companion. In such hierarchical triples, the inner binary can initially be wide enough for the BH to form without ever transferring mass to its companion. Furthermore, over long-term evolution, the tertiary can induce secular eccentric Kozai–Lidov (EKL) oscillations \citep[][]{lidov62,kozai62,Naoz16}, driving the inner binary to extreme eccentricities. Such high eccentricities drastically reduce the pericenter distance, allowing short-range forces, such as gravitational waves, tides, and magnetic braking, to shrink the orbit. For double-compact-object inner binaries, this triple-induced channel is efficient at producing highly eccentric GW sources \citep[][]{wen03,Hoang+18,Hamers+18,Stephan+19,Bub+20,Deme+20,Wang+21,Xuan23acc,Xuan+23b,Xuan24bkg,Xuan24parameter,xuan2025GC,stegmann2025}, while for BH–MS inner binaries it can efficiently form BH–LMXBs \citep[][]{Naoz16_LMXB,shariat2025triple}.

In fact, V404 Cygni, a prototypical BH–LMXB and one of the nearest known, was recently confirmed to host a tertiary companion \citep[][]{burdge2024}. Its properties are consistent with formation via the triple-induced high-eccentricity channel \citep[][]{shariat2025triple}, demonstrating that field BHs can reside in long-lived hierarchical triples and some are born with negligible natal kicks \citep[$v_{\rm k} \lesssim 5~{\rm km~s^{-1}}$][]{burdge2024,shariat2025triple}. Moreover, V404 Cygni likely evolved from an initially wide BH–MS binary \citep[][]{shariat2025triple}, and such a stellar population yields an even larger population of wide BH–WD inner binaries (separation $a_1\gsim10-100$~au) in triples. The long-lived, wide BH-WD inner binaries could later experience strong secular EKL oscillations and form UCXBs.

In this study, we investigate the dynamical evolution of detached BH–WD binaries in hierarchical triple systems. In particular, we model their evolution by incorporating eccentric Kozai–Lidov oscillations, general relativistic precession, gravitational wave emission, WD dynamical tides, and WD mass transfer. We then estimate the fraction of systems that evolve into BH–UCXBs, evaluate their expected observational signatures in both X-ray and GW bands, and predict the contribution of the triple channel to the UCXB population in the Milky Way and the Andromeda galaxy.

This letter is organized as follows. Section~\ref{subsec:initial} introduces the formation mechanism and initial population of detached BH–WDs with tertiaries. Section~\ref{subsec:dyntides} describes their general behavior in triple evolution and the treatment of WD dynamical tides. Section~\ref{subsec:masstransfer} focuses on the subsequent mass-transfer evolution of BH–WD systems that undergo strong orbital shrinkage and form UCXBs after the triple stage. We present the simulation results in Section~\ref{subsec:results}, evaluate the X-ray and GW detectabilities of BH–UCXBs in Section~\ref{subsec:detectability}, and outline the assumptions and limitations of our work in Section~\ref{subsec:caveats}. Finally, Section~\ref{sec:conclusion} discusses the results and their astrophysical implications.

\section{Simulation setup}
\label{sec:simulation}
\subsection{Generating the initial BH-WD population}
\label{subsec:initial}
In this work, we take the initial wide BH-WD population from the simulation results of \citet{shariat2025triple} (see their fig.4), and further evolve the hierarchical triple systems up to a Hubble time, or until mass transfer starts. For completeness, we briefly summarize their
population synthesis setup below, and discuss the resultant BH-WD population.

At t=0, the triples were initialized with all three stars on the zero-age main sequence (ZAMS): a primary of $m_1 = 22~{\rm M_\odot}$ \citep[expected to produce a $\sim10~{\rm M_\odot}$ black hole;][]{Sukhbold16}, {a secondary drawn from a uniform mass distribution between $m_2 \sim 1.2-2~{\rm M_\odot}$}, and a tertiary of $m_3 = 1.2~{\rm M_\odot}$, chosen to match the observed properties of V404 Cygni, the only known black hole in a triple \citep[][]{burdge2024}.

Inner and outer orbital periods were drawn from a log-uniform distribution between $0.1$ and $10^4$~yr \citep[e.g.,][]{Sana+12}.  
Eccentricities were sampled from a uniform distribution for the inner and outer orbits. Mutual inclinations were drawn from an isotropic distribution (uniform in $\cos i$), and spin–orbit angles chosen uniformly. This choice is consistent with observations indicating that wide triples have mutual inclinations consistent with an isotropic distribution \citep[e.g.,][]{Tokovinin22, Shariat25_10ktriples}.

Each set of initial conditions is required to satisfy both hierarchical and long-term dynamical stability criteria.  
For hierarchy, we use the octupole-level parameter $\epsilon$ \citep[e.g.,][]{Naoz+13},  
\begin{equation}\label{eq:eps_crit}
    \epsilon = \frac{a_1}{a_2} \frac{e_2}{1-e_2^2} < 0.1 \ ,
\end{equation}
where $a_1$ and $a_2$ are the inner and outer semi-major axes, and $e_2$ is the outer eccentricity. 

For long-term stability, we adopt the \citet{mardling01} criterion:
\begin{equation}\label{eq:MA_stability_crit}
    \frac{a_2}{a_1} > 2.8 \left(1 + \frac{m_3}{m_1 + m_2}\right)^{2/5}
    \frac{(1+e_2)^{2/5}}{(1-e_2)^{6/5}}
    \left(1 - \frac{0.3 i}{180^\circ}\right) \ ,
\end{equation}
where $m_1$ and $m_2$ are the inner binary masses, $m_3$ is the tertiary mass, and $i$ is the mutual inclination. Hereafter, we use $a$ to represent $a_1$.

Deviation from strict hierarchy does not necessarily lead to the immediate disruption of the system or the onset of dynamical instability \citep{Grishin17, Mushkin20, Bhaskar21, Toonen2022, Zhang23}. Nevertheless, recent observations indicate that virtually all field triples are hierarchical and stable according to the above criteria, reinforcing our choice to focus on hierarchically stable configurations \citep[][]{Shariat25_10ktriples}.

Using the above numerical setup, \citet{shariat2025triple} evolves a large population of 50,000 triples using detailed dynamical simulations, which includes the secular equations up to the octupole level of approximation \citep{Naoz+13,Naoz16}, and general relativity precession \citep[e.g.,][]{naoz13}. Prior to BH-WD formation, the simulations also included main-sequence stellar evolution using {\tt MESA} \citep{Paxton11} grids from the {\tt POSYDON} \citep[][]{Fragos23} framework. For main-sequence stars, {they} assumed equilibrium tides that change from convective to radiative depending on the stellar type.

Beyond producing a few hundred BH-LMXBs through the triple channel, this population produces an even larger number of detached BH+WD inner binaries ({a sample of 1242 systems}), which are the focus of this study. 
These BH-WD inner binaries in triples are all detached and have never transferred mass. Therefore, most of them are wide ($a_1\sim 10-4000$~au), with wide tertiaries companions ($a_2\sim 200-15000$~au), see e.g., Figure~\ref{fig:initial} in Appendix~\ref{sec:appendixA}. Notably, the inner binaries of these systems have presumably lost at least half of their initial mass, making their orbits at least twice as large as the initial ones.

\subsection{Evolution of BH-WDs in triples and the inclusion of dynamical tides}
\label{subsec:dyntides}

To evaluate the subsequent evolution of BH-WD systems, we further include the GW
emission \citep[e.g.,][]{Peters64,Zwick+20,Xuan+23b} and dynamical tides, in addition to the previous features of the triple simulation code.
Notably, when compact binaries have a pericenter passage time ranging from minutes to hours, dynamical tides excited in the WD component can play a crucial role in shaping the system’s physical properties, including the orbital separation, eccentricity, and WD spin. In this work, we adopt the models developed by \citet{Fuller2012, Vick2017, Su2022} and incorporate the dynamical tides into the long-term orbital evolution of hierarchical triple systems\footnote{{Note that here tidal dissipation in white dwarfs is assumed to be 100\% efficient.}}. We assume a representative mass of $m_1=10$~M$_{\odot},\, m_2=0.6$~M$_{\odot}$ for BH-WD binaries during this evolution stage \citep[e.g.,][]{corral16,Bahramian_2023,pelisoli2025observationaloverviewwhitedwarf}.

In particular, we follow eqs. 34-35 in \citet{Vick2017} to compute the angular momentum and energy transfer rate of dynamical tides in the inertial frame:
\begin{equation}
    \dot{J}_{\text {tide }}=T_0 \sum_{-\infty}^{\infty} F_{N 2}^2 \operatorname{sgn}\left(N \Omega-2 \Omega_{\mathrm{s}}\right) \hat{F}\left(\omega=\left|N \Omega-2 \Omega_{\mathrm{s}}\right|\right),
    \label{eq:Jdot}
\end{equation}
\begin{equation}
\begin{aligned}
& \dot{E}_{\text {tide, in }}= T_0\left[\left(\frac{W_{20}}{W_{22}}\right)^2 \sum_{N=1}^{\infty} N \Omega F_{N 0}^2 \hat{F}(\omega=|N \Omega|)\right. \\
& \left.+\frac{1}{2} \sum_{-\infty}^{\infty} N \Omega F_{N 2}^2 \operatorname{sgn}\left(N \Omega-2 \Omega_{\mathrm{s}}\right) \hat{F}\left(\omega=\left|N \Omega-2 \Omega_{\mathrm{s}}\right|\right)\right]\,,
\end{aligned}
    \label{eq:Edot}
\end{equation}
where $T_0 \equiv G m_1^2 R_2^5/a^6$ is a function of WD radius\footnote{The WD mass {stays} unchanged during the triple evolution, thus, for simplicity, we assume a fixed $R_2\sim 5.3\times 10^{-5}$~au when computing the dynamical tides.} $ R_2$,$W_{20}=\sqrt{\pi/5}$, $W_{22}=\sqrt{3\pi/10}$; $\Omega$ is the orbital frequency of the binary, $\Omega_s$ is the WD spin angular frequency, and $\hat{F}(\omega)$ is the dimensionless tidal torque of the WD\footnote{$\hat{F}(\omega)\sim 150\omega^5$ for a $T_2\sim5000$~K WD, as a function of tidal frequency, $\omega$ in $G=M=R=1$ unit, see e.g., fig.6 in \citet{Vick2017}.}. Hereafter, we use $\dot{E}_{\rm tide}$ to represent $\dot{E}_{\rm tide,in}$ for simplicity.

In Equations~(\ref{eq:Jdot}) and (\ref{eq:Edot}), the summing of integer number $N$ reflects the multiple harmonics of dynamical tides for a general eccentric orbit, and $F_{Nm}$ is the Hansen coefficient \citep{Murray1999,Storch_2013}:
\begin{equation}
    F_{N m}=\frac{1}{\pi} \int_0^\pi \frac{\cos [N(E-e \sin E)-m f(E)]}{(1-e \cos E)^2} \mathrm{~d} E \ ,
\end{equation}
in which $e$ is the eccentricity of the BH-WD binary; $f(E)$ is the true anomaly, as a function of eccentric anomaly $E$.

We note that, the Hansen coefficients $F_{N 2}$ and $F_{N 0}$ can be evaluated using analytical approximations \citep{Su2022}:
\begin{equation}
F_{N 2} \approx \begin{cases}C_2 N^p e^{-N / \eta_2} & N \geq 0 \\ 0 & N<0\end{cases}
\label{eq:Fn2}
\end{equation}
and
\begin{equation}
F_{N 0}=C_0 e^{-|N| / \eta_0}\,,
\label{eq:Fn0}
\end{equation}
in which $C_0, C_2, p$, and $\eta_0, \eta_2$ are fitting coefficients that can be expressed as a function of binary eccentricity (see eqs. (20) - (31) in \citet{Su2022}). To accelerate the numerical simulation, we further employ empirical relations to analytically fit the results of Equations (\ref{eq:Jdot})–(\ref{eq:Fn0}); see Appendix~\ref{sec:appendixB} for details.

Furthermore, the orbital evolution of the binary can be related to the energy and angular momentum transfer rate \citep{Su2022}:
\begin{equation}
    \frac{\dot{a}_{\rm tide}}{a}=-\frac{2 a \dot{E}_{\mathrm{tide}}}{G m_1 m_2},
\end{equation}
\begin{equation}
\frac{e\dot{e}_{\rm tide} }{1-e^2}=-\frac{a \dot{E}_{\mathrm{tide}}}{G m_1 m_2}+\frac{T}{J_{\mathrm{orb}}},
\end{equation}
where $J_{\text {orb }}=m_1 m_2\left[G a\left(1-e^2\right) /\left(m_1+m_2\right)\right]^{1 / 2}$ is the orbital angular momentum and $T$ is tidal torque, which equals the angular momentum transfer rate $\dot{J}_{\rm tide}$. Note that here we only show the time derivative of orbital separation and eccentricity ($\dot{a}_{\rm tide}$ and $\dot{e}_{\rm tide}$) contributed by dynamical tides. In a realistic simulation, there will be additional terms from GW radiation \citep[][]{Peters64} and triple interaction \citep[][]{Naoz16} that contribute to the final result of orbit evolution.

Additionally, as can be seen from Equations~(\ref{eq:Jdot}) and (\ref{eq:Edot}), $\Omega_s$ can significantly affect the tidal torque. Therefore, it is necessary to include the spin evolution in the simulation. Specifically, we compute its time derivative, $\dot{\Omega}_s$, assuming that the WD rotates rigidly:
\begin{equation}
\dot{\Omega}_{\mathrm{s}}=\frac{T}{k m_2 R_2^2}
\label{eq:Osdot}
\end{equation}
where $k m_2 R_2^2$ is the moment of inertia of the WD. For simplicity, we assume that $k=0.2$, and initialize the spin of the WD to zero, i.e., $\Omega_s=0$. We further assume that the spin aligns with the orbital angular momentum. This is because the spin of WD is only excited after dynamical tides become strong, which have shrunk the orbit, suppressed the EKL oscillation, and frozen the orientation of the system.


\begin{figure}
    \centering
    \includegraphics[width=3.5in]{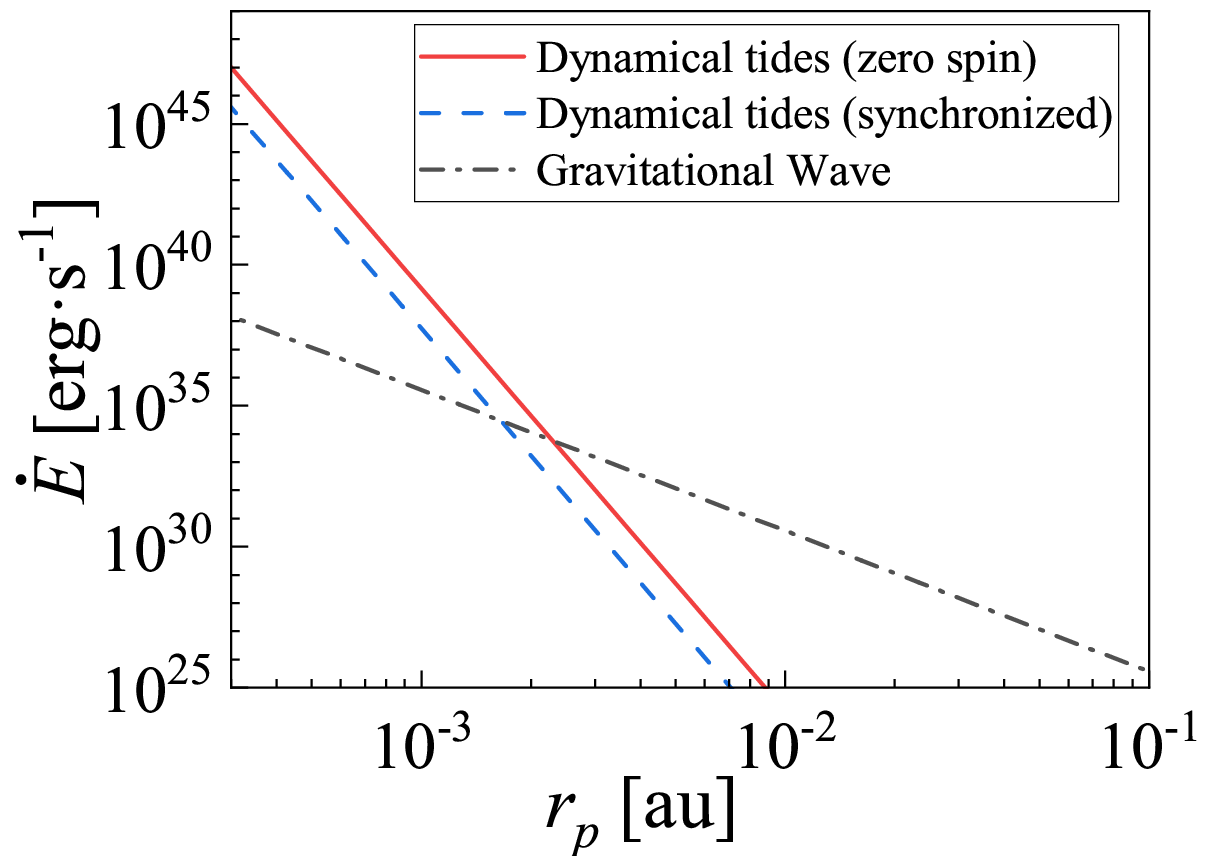}
    \caption{{\bf Comparison between the energy transfer rate of dynamical tides and GW radiation, for a highly eccentric BH–WD system.} Here we consider a $10$–$0.6~\rm M_{\odot}$ BH–WD binary with $e=0.95$, WD radius $R_2=5.3\times 10^{-5}$~au, and WD dimensionless tidal torque $\hat{F}(\omega)\sim 150\omega^5$. The energy transfer rates from WD dynamical tides and from GW radiation are plotted as functions of the binary’s pericenter distance $r_p$. We show two cases for dynamical tides: the zero-spin case (red solid line, fixing $\Omega_s=0$) and the synchronized-spin case (blue dashed line, when $\Omega_s$ synchronizes with pericenter frequency and satisfies $\dot{J}_{\rm tide}=0$). The energy loss due to GW radiation is shown as the grey dash-dotted line. We note that when eccentricity is nonzero, there will be a residual tidal heating rate (blue dashed line) even when there is no net torque on the WD. This rate will vanish in the circular limit.} 
    \label{fig:tides}
\end{figure}

During the long-term orbital evolution, the WD spin tends to pseudo-synchronize with the orbital pericenter frequency, $\Omega_p \sim \Omega (1-e)^{-3/2}$ \citep{Vick2017}. In this case, the angular momentum transfer rate will vanish, yielding zero net torque on the WD (see Equation~(\ref{eq:Jdot})). However, if the orbital eccentricity is non-zero, it can always excite multiple harmonics of dynamical tides in the WD, which creates a residual tidal heating rate. In other words, even if $\dot{J}_{\rm tide}=0$, $\dot{E}_{\rm tide}$ in Equation~(\ref{eq:Edot}) keeps non-zero. For systems in our simulation, Equation~(\ref{eq:Edot}) will only vanish when both the spin is fully synchronized and the orbit gets fully circularized.

\begin{figure*}
    \centering
    \includegraphics[width=6in]{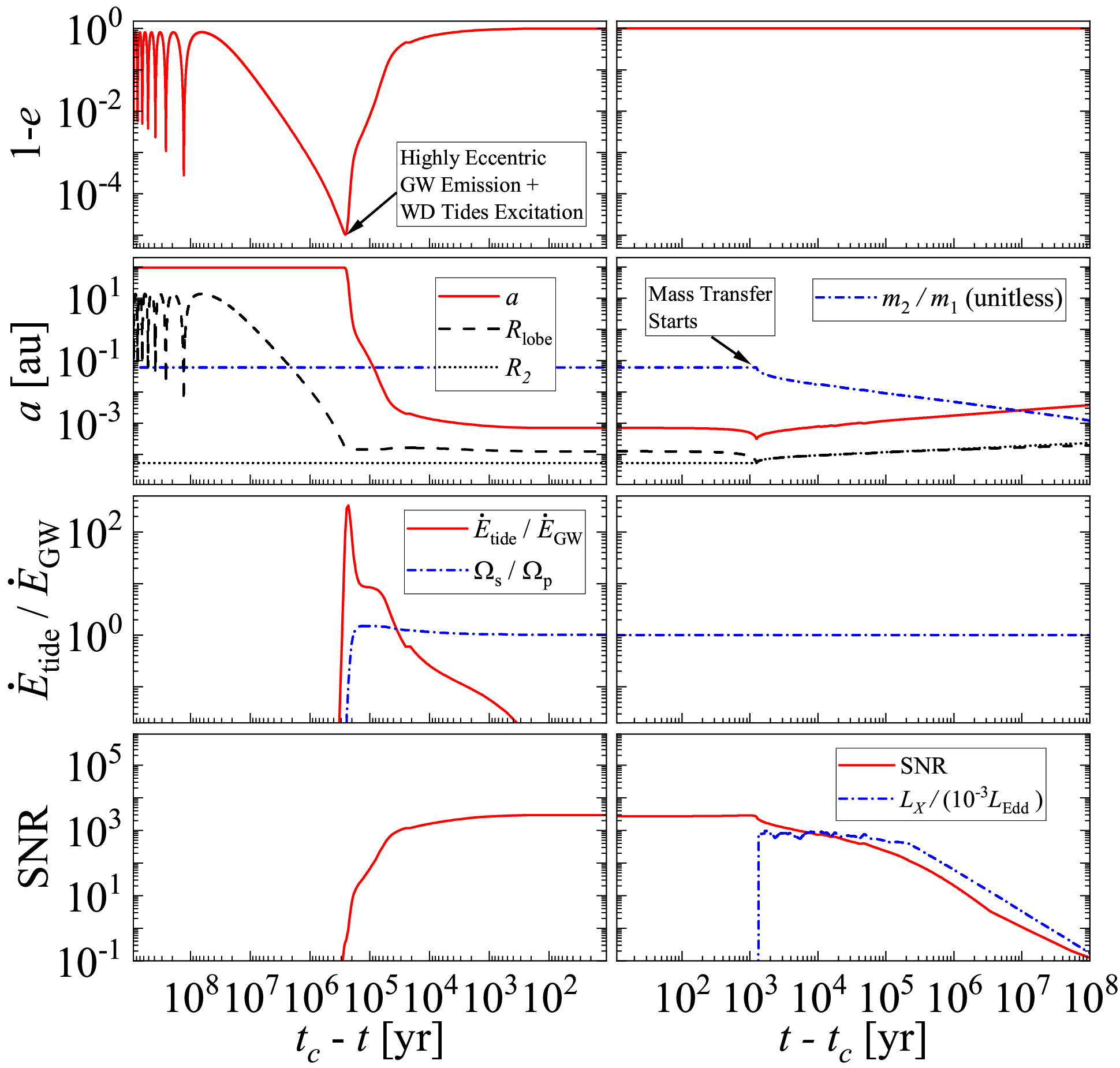}
    \caption{{\bf{Time evolution of an example BH-WD system that undergoes strong EKL oscillations during the dynamical formation stage (left column) and transitions into a mass-transferring UCXB (right column).}} Here we show the time evolution of a representative BH-WD system from our simulation, with initial mass $m_1=10$~$\rm M_{\odot}$, $m_2=0.6$~$\rm M_{\odot}$, orbit separation $a_0=95.3$~au, eccentricity $e_0=0.199$. This binary is orbited by a tertiary companion with mass $m_3=0.836$~$\rm M_{\odot}$, outer orbit separation $a_2=4997$~au, eccentricity $e_2=0.602$, and mutual inclination $i = 91.8^\circ$ to the inner binary. We define $t_c$ as the time when the BH–WD binary becomes fully circularized ($e < 0.01$). The system's evolution during the dynamical formation stage is shown in the left column as a function of $t_c - t$ (log-scale), while its post-circularization evolution is shown in the right column as a function of $t - t_c$. The first row shows the inner binary's orbital eccentricity, as $1-e$. The second row shows the inner orbital separation (red solid line, $a$), the WD–BH mass ratio (blue dash-dotted line, $m_2/m_1$), the Roche lobe radius (black dashed line, $R_{\rm lobe}$, see, e.g., eq.28 in \citet{Ye_2023}), and the WD radius (black dotted line, $R_2$). The third row plots two ratios: the tidal energy dissipation rate (red solid line, see Equation~(\ref{eq:Edot})) to the GW radiation power (see, e.g., eq. 5.4 in \citet{Peters64}), and the WD spin angular frequency $\Omega_s$ to the binary’s pericenter frequency $\Omega_p = \Omega (1 - e)^{-3/2}$ (blue dash-dotted line). The bottom row shows the expected gravitational wave SNR at distance $D_l=8$~kpc (red solid line) for a 4-year LISA observation, and the binary’s X-ray luminosity $L_X$ (blue dash-dotted line), normalized by $0.1\%$ of the BH Eddington luminosity, $L_0=10^{-3}L_{\rm Edd} \sim 1.26\times 10^{36}$erg~s$^{-1}$.}
    \label{fig:eg1}
\end{figure*}

Figure~\ref{fig:tides} shows the energy transfer rate of dynamical tide (red solid line, Equation~(\ref{eq:Edot})), compared with the power of GW radiation \citep[grey dash-dotted line,][]{Peters64}, for a highly eccentric BH-WD binary. As can be seen in the figure, the effect of dynamical tides strongly depends on the pericenter distance. Notably, it can well-exceed the GW radiation power when $r_p\lesssim10^{-3}$~au. This means WD tides can effectively shrink the orbit during the highly eccentric pericenter passage and help with the formation of UCXBs. Furthermore, even when the WD spin is synchronized with the orbit (blue dashed line, $\Omega_s\sim \Omega_p$), the residual tidal heating is still significant, because of the eccentricity. This means the dynamical tide can keep dissipating energy efficiently, until the binary circularizes and $\dot{E}_{\rm tide}$ vanishes.

We further illustrate the time evolution of a representative BH–WD system in Figure~\ref{fig:eg1}, with the left column showing its dynamical formation stage. Initially, the inner binary undergoes significant EKL effect because of the distant tertiary companion, which is reflected in the top-left panel in the form of eccentricity oscillation. At the time $t_c - t \sim 0.3$~Myr, the inner binary reaches extreme eccentricity ($1 - e \sim 10^{-5}$), driving the pericenter distance down to $\sim 10^{-3}$~au and triggering strong dynamical tides. In particular, the dynamical tide is excited under conditions of high eccentricity and negligible WD spin, which results in a sharp peak in the ratio of tidal dissipation to GW radiation power (see third row, left panel). As the system evolves, the WD spin synchronizes with the orbital pericenter motion over a short timescale ($\sim 10^5$~yr, see the blue dash-dotted line). This synchronization occurs along with the significant orbital decay and circularization, which eventually suppresses the dynamical tides and leads to the decoupling of the inner binary from the influence of the tertiary companion. By the time the orbital separation $a$ shrinks to $\sim 10^{-3}$~au, systems typically exhibit nearly circular orbits and synchronized WD spins (see, e.g., when $t_c-t\lesssim 10^2$~yr).

\subsection{Evolution of mass-transfer BH-WDs}
\label{subsec:masstransfer} 

Following the highly eccentric dynamical evolution (see Section~\ref{subsec:dyntides}), the BH-WD system's orbit can further shrink and circularize under the influence of GW emission and tidal dissipation, until the Roche lobe of the white dwarf component is smaller than its volume. At this stage, the white dwarf starts to lose surface matter, transferring mass to its BH companion. This process has been studied in depth and is expected to cause X-ray emission, making the system observable as a UCXB. \citep[e.g.,][]{Deloye_2003,van_Haaften_2012,Heinke_2013,Sengar_2017,Bobrick2017,Church2017,Chen_2020UCXB,Suvorov_2021,Qin_2023,chen2025newpotentialultracompactxray}

To estimate the orbital evolution of a UCXB system, we assume the binary has decoupled from the tertiary and evolves in isolation, with the orbit fully circularized and the WD spin synchronized\footnote{When synchronized with the orbit, the WD's spin angular momentum is usually much smaller than the orbital angular momentum (e.g., $J_{\rm spin}/J_{\rm orb}\lesssim 10^{-4}$ for orbital frequency lower than 10mHz). Therefore, we neglect the term caused by spin angular momentum transfer.} (see the discussion in Section~\ref{subsec:dyntides}). For simplicity, we also assume that the accretion is stable, with mass lost by the donor either accreted or leaving the system as unbound material. 
Therefore, the change of total orbital angular momentum is given by \citep[see, e.g., eq.2 in][]{van_Haaften_2012}\footnote{As a proof of concept, we adopt this simplified model which ignores the feedback from the disk. For a more detailed analysis, see, e.g., \citet{Deloye_2003,van_Haaften_2012,Sengar_2017}.  }:
\begin{equation}
\dot{J}_{\mathrm{orb}}=\dot{J}_{\mathrm{GW}}+\dot{J}_{\mathrm{eject}} .
\label{eq:Jflow}
\end{equation}
where $\dot{J}_{\rm GW}$ is the angular momentum loss caused by the binary's gravitational wave radiation \citep[eq 5.5 in][]{Peters64}, and $\dot{J}_{\mathrm{eject}}$ is the angular momentum loss caused by the ejected matter \citep[see, e.g., eq.18 in][]{van_Haaften_2012}:
\begin{equation}
    \frac{\dot{J}_{\mathrm{eject}}}{J_{\rm orb}}=q\frac{\dot{M}_{\rm tot}}{M_{\rm tot}} .
\label{eq:jeject}
\end{equation}
where $q=m_2/m_1$ is the mass ratio of the binary, $M_{\rm tot}=m_1+m_2$, and $\dot{M}_{\rm tot}$ is the mass loss rate of the system.

On the other hand, from the functional form of the orbital angular momentum $J_{\rm orb}$, we have \citep[see, e.g., eq.23 in][]{van_Haaften_2012}:
\begin{equation}
\frac{\dot{J}_{\rm orb}}{J_{\rm orb}}=\frac{\dot{m}_1}{m_1}+\frac{\dot{m}_2}{m_2}-\frac{1}{2} \frac{\dot{M}_{\rm tot}}{{M}_{\rm tot}}+\frac{1}{2} \frac{\dot{a}}{a} .
\label{eq:Jdot_masstransfer}
\end{equation}
where $\dot{m}_1$, $\dot{m}_2$ are the mass transfer rate of BH (positive) and WD (negative), respectively. Define $\eta$ as the accretion efficiency, we have $\dot{m}_1=-\eta \dot{m}_2$, and $\dot{M}_{\rm tot}=\dot{m}_1+\dot{m}_2=(1-\eta)\dot{m}_2$. 

The change of orbital separation can be estimated by combining Equations~(\ref{eq:Jflow}) and (\ref{eq:Jdot_masstransfer}), then inserting Equation~(\ref{eq:jeject}):
\begin{equation}
\frac{1}{2} \frac{\dot{a}}{a}
=
\frac{\dot{J}_{\mathrm{GW}}}{J_{\mathrm{orb}}}
-
\frac{\dot{m}_2}{m_2}
\left[
1-  q + (1 - \eta)\frac{q}{2(q + 1)}
\right] .
\label{eq:adot}
\end{equation}
Notably, the coefficient of $\dot{m}_2/{m_2}$ in the right side of Equation~(\ref{eq:adot}) is always negative, provided that $q<1$ and $\eta<1$. In other words, stable mass transfer from WD to BH ($\dot{m}_2<0$) tends to widen the orbit ($\dot{a}>0$), stopping the binary from orbit shrinkage by GW radiation, and making them long-living UCXB sources.

Furthermore, we adopt the result of \citet{van_Haaften_2012}, taking $\dot{m}_2$, and $\eta$ as functions of the UCXB orbital parameters from their figs.~1 and 4. Consequently, all the quantities in Equation~(\ref{eq:adot}) can be expressed in terms of $m_1$, $m_2$, and $a$, and the system's evolution can be determined by integrating the differential equations of $dm_1/dt,\,dm_2/dt$, and $da/dt$. 

We evolve the BH-WD systems until the WD mass is too light (specifically $m_2 \sim 0.01$–$0.02$~M$_{\odot}$, which may result in disrupting the WD), and the orbit is too wide to maintain a stable mass transfer. Notably, the final WD mass puts a constraint on the lifetime of a UCXB system, yet this value remains highly uncertain. In particular, previous studies suggest that the accretion disk in a BH–WD UCXB becomes thermally unstable when the donor mass decreases to $m_2 \sim 0.02$~M$_{\odot}$ at an orbital period of $P_{\rm orb} \sim 28$ minutes \citep[orbital frequency $f_{\rm orb}\sim 0.6$~mHz, see][]{van_Haaften_2012}. This WD mass value is supported by recent observations that have detected BH-WD UCXBs with donor masses $m_2 \sim 0.01-0.016$~M$_{\odot}$ \citep[see, e.g.,][]{Tudor2018}. However, \citet{van_Haaften_2012} also estimates that BH–WD systems may evolve to orbital periods as long as 110 minutes with a final donor mass $m_2 \sim 0.0026$~M$_{\odot}$, provided the system is not disrupted. 

Here, we assume that the UCXB phase terminates when the WD mass decreases to $m_2 \sim 0.01$--0.02~M$_\odot$ {(representing an estimated range of final WD mass, 
based on discussions in \citet{van_Haaften_2012} and observed UCXB donor masses). 
Substituting this range into Equation~(\ref{eq:adot}) gives a UCXB lifetime of $\tau \sim 20$--180~Myr for a 10--0.6~M$_\odot$ system.} 
We note, however, that if the final donor mass can indeed extend to extremely low values
($\sim 0.0026$~M$_\odot$), the UCXB lifetime could extend up to $\sim 10$~Gyr.  

In the right column of Figure~\ref{fig:eg1}, we show the mass-transfer evolution of a representative BH–WD system, following its earlier dynamical evolution shown in the left column (see Section~\ref{subsec:dyntides}). As illustrated in the second-row panels, the mass transfer starts at $t-t_c\sim 6\times10^3$~yr, when the orbital separation drops to $\sim 3\times 10^{-4}$~au (red solid line) and the WD radius (black dotted line) fills its Roche lobe (black dashed line). Following the onset of mass transfer, the WD begins to lose mass, which increases the orbital separation $a$ (see Equation~(\ref{eq:adot})) and causes the expansion of the WD radius $R_2$. Consequently, the Roche lobe radius increases along with the WD radius, leading to a wider orbit, a decreasing mass ratio $m_2/m_1$ (see the blue dash-dotted line), and a gradual decrease in gravitational wave emission (see the red line in the last row).

Furthermore, we compute the binary's X-ray luminosity using the accretion rate of the BH, assuming that $L_X \sim 0.1 \dot{m}_1 c^2$ (see the blue dash-dotted line of the last row). Notably, the X-ray luminosity of simulated UCXBs typically lies between $10^{35}-10^{38}$~erg~s$^{-1}$. Assuming a luminosity distance of $ D_l\sim 20$~kpc, these systems can be observed with an X-ray flux of $F_X=L_X/(4\pi D_l^2)\gtrsim 10^{-12}$~erg~cm$^{-2}$~s$^{-1}$, which is exceeding the detection sensitivity of most present-day X-ray telescopes ($\sim 10^{-14}$~erg cm$^{-2}$ s$^{-1}$, see, e.g., \citet{Kim_2007}). In other words, once a mass-transferring BH-WD system is formed, its X-ray signal is well detectable in the Milky Way throughout the stable mass-transfer lifetime\footnote{Assuming the WD final mass is as low as $\sim 0.0026$~M$_{\odot}$ and the system evolves up to 10~Gyr, its X-ray flux is still beyond $F_X\gtrsim10^{-14}$~erg~cm$^{-2}$~s$^{-1}$ at 20~kpc.}. Furthermore, the detection of BH-UCXBs can be well extended to other galaxies, up to a luminosity distance of $\sim 10-100$~Mpc, see, e.g., Figure~\ref{fig:dmax_xray} in Appendix~\ref{sec:appendixC}.


\section{Results and discussion}
\label{sec:result}
\begin{figure*}
    \centering
    \includegraphics[width=7in]{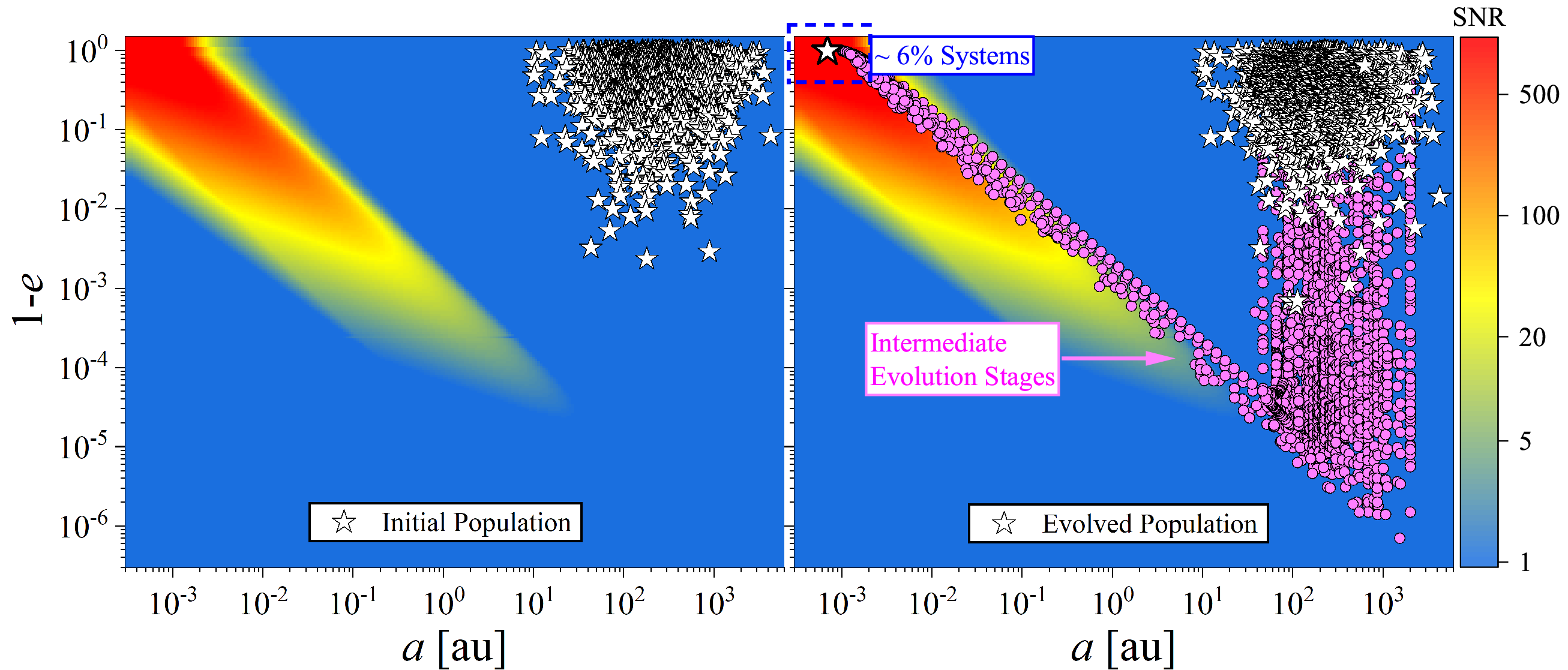}
    \caption{{\bf The population of BH-WD systems from simulated Milky Way Galactic field, and their estimated GW SNR as a function of semi-major axis and eccentricity (for a 4-yr LISA observation). } Here, we plot the semi-major axis, $a$, and eccentricity, as $1-e$, of each simulated BH-WD system (see the white stars). The {\it Left Panel} shows the initial conditions, while the {\it Right Panel} represents the final state of each system, either when its dynamical evolution ends at the Hubble time or when the orbit separation shrinks to $\lesssim10^{-3}$~au and triggers the mass transfer. The background color maps the binary's expected gravitational wave SNR, assuming a fixed distance $D_l=8$~kpc and a 4-yr LISA observation. In the {\it Right Panel}, we also plot intermediate evolutionary stages (pink dots) for binaries that eventually evolve into UCXBs. These dots represent the evolution track of UCXBs formed via the wide triple channel, and highlight the region of parameter space where they may become detectable. } 
    \label{fig:population}
\end{figure*}

\subsection{Triple Path of UCXB Formation}
\label{subsec:results}
Following the method described in Section~\ref{sec:simulation}, we adopt {a sample of 1242 detached BH-WD systems with stellar-mass tertiary companions} (see, e.g., the initial condition in Appendix~\ref{sec:appendixA}, Figure~\ref{fig:initial}). These triples have survived the main sequence evolution in the previous study \citep{shariat2025triple}, and meet the criteria of hierarchy and long-term stability. These systems' initial configurations are depicted in the ($1-e$) - $a$ parameter space in Figure \ref{fig:population}, left panel. After evolving them, we identify 78 out of 1242 that end up as ultracompact X-ray binaries, corresponding to a formation efficiency of 
$f_{\rm UCXB}=6.3\%$. The evolution trajectories of these systems are shown as pink points in the right panel of Figure~\ref{fig:population}. 


As shown by the figure, during the dynamical formation stages, these systems are perturbed by the EKL effect and excited to extreme eccentricities, $(1-e)\sim10^{-7}-10^{-5}$. {However, the rapid circularization due to dynamical tides typically causes the binary to first evolve into circular configurations, rather than becoming even more eccentric and being tidally disrupted.} Furthermore, the combination of high eccentricity and small pericenter distance leads to rapid orbital decay driven by GW emission and WD dynamical tides, with an average GW detectable lifetime $\tau_{\rm GW,\,dyn}\sim 0.25$~Myr of this stage (assuming random distance in MW, 4 yr LISA detection, see Section~\ref{subsec:detectability} for details). By the end of this stage, the orbit of these BH-WDs typically shrinks to $\sim 10^{-3}$~au with residual eccentricity $\lesssim 0.01$, and the WD spin frequencies are mostly synchronized with their orbit frequency.

Following the orbital shrinkage and the decoupling from EKL oscillations, the BH-WD systems start mass transfer and X-ray emission (at $a\sim 3\times10^{-4}$~au). During this stage, the binary undergoes long-term evolution due to the combined effect of GW radiation and mass transfer, resulting in a gradual widening of the orbit (see, e.g., Equation~(\ref{eq:adot}) and the right column of Figure~\ref{fig:eg1}). From our simulations, we infer an averaged GW detectable lifetime of UCXB stage $\tau_{\rm GW,\,UCXB}\sim 3.6$~Myr, and X-ray detectable time $\tau_{\rm UCXB}\sim 20-180$~Myr in the MW, depending on the assumed final WD mass ($0.02-0.01$~M$_{\odot}$) when the mass transfer becomes unstable (see Section~\ref{subsec:masstransfer}).

\subsection{ UCXB detectability and expected population}
\label{subsec:detectability}
To estimate the number of detectable BH-WD systems in GW and X-ray observations, we need to take into account both their formation rate and detectable time. Particularly, here we assume a constant star formation rate (SFR), i.e., a continuous birth and death of compact object binaries in the Milky Way, keeping the total number of systems unchanged. Thus, the expected number of systems in a given stage is simply achieved by taking the UCXB rate, $\Gamma_{\text {UCXB}}$, and multiplying by the average detectable lifetime, $\tau$. Namely: 
\begin{equation}
   N=\Gamma_{\text {UCXB}}\,\tau \ .
    \label{eq:steadystate}
\end{equation}
To estimate the formation rate of UCXBs in the Milky Way, we adopt the following relation:
{\begin{equation}
\begin{aligned}
\Gamma_{\text {UCXB}} = & \mathrm{SFR} \times f_{\text {triple}} \times f_{m_1,m_2,m_3} \\
& \times f_{\text {no kick}}\times f_{\text {BH-WD triple}}   \times f_{\rm UCXB},
\label{eq:formation_rate}
\end{aligned}
\end{equation}}
where $\rm SFR\sim 2 ~yr^{-1}$, which roughly corresponds to a constant star formation rate of $1~{\rm M_\odot~yr^{-1}}$ with a Kroupa IMF \citep[][]{Kroupa2001}. {$f_{\text {triple}} = 0.11$ is the fraction {\it of all stars formed} that reside in triples \citep{Shariat25_10ktriples} and $f_{m_1,m_2,m_3}$ is the fraction of triples with $20<m_1/{\rm M_\odot}<40$, $0.8<m_2/{\rm M_\odot}<8$, and $0.5<m_3/{\rm M_\odot}<8$, chosen such that the primary becomes a black hole, the secondary becomes a white dwarf within a Hubble time, and the tertiary does not become a neutron star/BH. We derive these values by generating a mock stellar population with singles, binaries, and triples following \citet{Shariat25_10ktriples}. Their publicly available prescription\footnote{\url{https://github.com/cheyanneshariat/gaia_triples}} samples inner binary parameters following the multidimensional distributions of \citet{Moe+17}
and tertiary masses from a $q^{-1.4}$ power law distribution. \citet{Shariat25_10ktriples} calibrate this procedure to maintain empirically constrained multiplicity statistics \citep[e.g.,][]{Moe+17,Tokovinin14,Winters19,Moe21} and to reproduce observations of the {\it Gaia} resolved triple population. This provides $f_{m_1,m_2,m_3}=2\times10^{-3}$ and $f_{\rm triple} = 0.11$.} Additionally, $f_{\text {BH-WD triple}} \sim 5.49\%$ is the fraction of triples that survive the main sequence evolution and the inner binary evolves into a detached BH-WD system \citep[see table 2 in][]{shariat2025triple}. $f_{\rm UCXB}\sim 6.3\%$ is the fraction of detached BH-WDs with a tertiary companion that undergo strong EKL oscillation and end up being a UCXB source\footnote{Note that $f_{\rm UCXB}$ is inferred from the dynamical simulation results of this paper, see Section~\ref{sec:result}}.

{$f_{\rm no, kick}$ denotes the fraction of triples that experience negligible BH natal kicks. We introduce this factor because the estimate of $f_{\text{BH–WD triple}}$ in \citet{shariat2025triple} assumes no natal kicks, whereas in reality some triples can receive strong kicks, become unbound, and need to be excluded from the rate estimate (see Section~\ref{subsec:caveats} for details). The strength and prevalence of BH natal kicks remain highly uncertain. Observations of BH–LMXBs indicate that some systems form with little to no kick while others experience strong ones
\citep[e.g.,][though selection biases favor weak kicks]{nagarajan2025kick}. Notably, the only confirmed BH–LMXB in a triple, V404 Cygni, also appears to have formed with a negligible kick \citep[$v_{\rm k}\lesssim5~{\rm km~s^{-1}}$;][]{burdge2024,shariat2025triple}. For conservative purposes, we therefore adopt $f_{\rm no, kick}\sim 10\%$.
Incorporating this fraction into Equation~(\ref{eq:formation_rate}) gives a total BH–UCXB formation rate of $\Gamma_{\rm UCXB} \sim 1.5 \times 10^{-7}~\mathrm{yr^{-1}}$ in the Milky Way.}

Thus, combining Equations~(\ref{eq:steadystate}) and (\ref{eq:formation_rate}), we have:
\begin{equation}
\begin{aligned}
    N_{\rm UCXB} & =\Gamma_{\rm UCXB}\times \tau_{\rm UCXB}\\ &\sim 1.5\times 10^{-7} {\rm yr}^{-1}\times \tau_{\rm UCXB}, \\ &\sim 3-27 \ ,
    \label{eq:NUCXB}
\end{aligned}
\end{equation}
in which $\tau_{\rm UCXB}$ is the average X-ray detectable time for a UCXB in the Milky Way.

In the last transition of Equation~(\ref{eq:NUCXB}), we adopt a lifetime of $\tau_{\rm UCXB} \sim 20-180$~Myr. This value is motivated by the fact that the X-ray luminosity of a UCXB system typically exceeds $\sim 10^{34}$erg~s$^{-1}$ throughout its evolution (see discussion in Section~\ref{subsec:masstransfer}). Consequently, such systems are well detectable within the Milky Way, with X-ray fluxes $\gtrsim 10^{-12}$erg~cm$^{-2}$~s$^{-1}$ above the sensitivity thresholds of most present-day X-ray observatories. Therefore, the average detectable time of a UCXB in the Galaxy can be approximated by its mass-transfer lifetime ($\sim 10^{7}$–$10^{8}$~yr; it may be as long as 10~Gyr in the most optimistic case, see Section~\ref{subsec:masstransfer}). Such a timescale is primarily determined by the orbital separation and initial mass at the onset of mass transfer, thus is nearly identical across the BH–WD systems considered in this paper.

On the other hand, however, the gravitational wave signals from UCXBs are not always detectable in the Milky Way, as can be seen from the signal-to-noise ratio (SNR) in the last row of Figure~\ref{fig:eg1}. Therefore, we further take into account the distance and SNR evolution of UCXBs to estimate their average GW detectable time. 

Specifically, we randomly generate BH-WD systems in the disk of MW, following the density profile of:
\begin{equation}
\rho(R, z) = \rho_0 \, \exp\left(-\frac{R}{R_d}\right) \, \exp\left(-\frac{|z|}{z_d}\right)
\end{equation}
where $\rho_0$ is a constant, $R$ is the galactocentric cylindrical radius, $z$ is the vertical height above the galactic plane, $R_d\approx {2.6}{\rm kpc}$ is the radial scale length, and $z_d\approx {0.3}{\rm kpc}$ is the vertical scale height \citep{Juri2008MW}.

For each generated system, we look into its time evolution, computing the GW SNR at all evolution stages \citep{Xuan+23b}:
\begin{equation}
{\rm SNR} 
\sim \frac{h_{\rm burst}}{\sqrt{S_n\left(f_{\rm burst }\right)}} \sqrt{T_{\rm o b s}\left(1-e\right)^{3 / 2}}, 
\label{eq:snrnew}
\end{equation}
where $ h_{\rm burst} \sim \sqrt{\frac{32}{5}}\frac{m_1m_2}{D_la(1-e)}$ is the peak GW strain amplitude, $f_{\rm burst} = f_{\rm orb} (1+e)^{1/2} (1-e)^{-3/2}$ is the peak GW frequency, $S_{\mathrm{n}}(f)$ is the
spectral noise density of LISA evaluated at GW frequency $f$
\citep[][]{2016PhRvD..93b4003K,Robson+19}, and $T_{\rm obs}=4$~yr is the LISA observation time. We then count the total amount of time each system spent observable in the LISA band (SNR larger than 8 for 4-yr observation), and estimate the average GW detectable time of BH-UCXB systems in the Milky Way. Our estimation yields $\tau_{\rm GW,\, dyn}\sim 0.25$~Myr for the eccentric dynamical formation stage, and $\tau_{\rm GW,\, UCXB}\sim 3.6$~Myr for the circular mass transfer stage (see appendix A1 of \citet{Xuan+23b}, for a similar method).

Therefore, our analysis yields the following number of GW-detectable sources\footnote{The expected GW source number can increase (to $\sim 2$) if LISA mission is extended to 10 years.}:
\begin{equation}
\begin{aligned}
    N_{\rm GW, \,UCXB} & =\Gamma_{\rm UCXB}\times (\tau_{\rm GW,\, dyn}+\tau_{\rm GW,\,UCXB})\\ &\sim 1.5\times 10^{-7} {\rm yr}^{-1}\times 4\,{\text{Myr}}, \\ &\sim 1 \ ,
    \label{eq:NUCXBGW}
\end{aligned}
\end{equation}
where we adopted a similar approach to the calculation presented in Equation~(\ref{eq:NUCXB}), only here we consider the overall GW detectable lifetime. Notably, since the average GW-detectable lifetime during the mass transfer ($3.6$~Myr) exceeds the dynamical formation ($0.25$~Myr), we expect the BH-WD GW sources to most likely appear as circular, mass-transferring systems. 

We further estimate the number of extragalactic sources contributed by this triple channel (see, e.g., their average detectable time as a function of distance in Appendix~\ref{sec:appendixC}). Using the Andromeda galaxy as an example, assuming a star formation rate of $0.4$~M$_{\odot}$ yr$^{-1}$ \citep{Rahmani_2016}, we find that Andromeda could host $\sim$1–5 UCXBs formed through this channel. However, due to the large distance, only a negligible fraction of them ($\sim1\%$) are expected to be detectable GW sources. These systems exhibit observational features consistent with some recently detected periodic X-ray sources \citep[e.g.,][]{zhang2024chandrasearchperiodicxray}, and may offer a potential explanation for the short-period low-mass X-ray binaries observed in M31.

\subsection{Discussion}
\label{subsec:caveats}
In this study, we specifically focus on the UCXB formation channel through isolated field triples. However, we note that other potential pathways, such as the mass transfer evolution of BH–MS binaries \citep[e.g.,][]{Qin_2023}, or dynamical formation through frequent interactions in globular clusters \citep[e.g.,][]{Ivanova_2010,Ivanova_2017}, may also contribute substantially to the Galactic UCXB population. 

While BH–UCXBs formed through different channels may appear observationally similar, as circular, mass-transferring BH–WD binaries, those formed through the wide triple channel could potentially be differentiated by identifying a distant tertiary companion ($a_2\sim 500-15000$~au) using photometric and/or astrometric observations. Additionally, LISA may be able to detect highly eccentric BH–WD progenitors in the Milky Way before they begin mass transfer, presenting a promising opportunity to distinguish between these formation channels. {We note that EKL is not the only mechanism that can drive eccentricity in BH–WD systems. For example, mass-loss-induced kicks \citep[][]{Hills1983,Kalogera_2000,Lu_2019,Shariat2023}, fly-by interactions \citep[][]{Michaely+20,Michaely+22}, and collisions \citep[][]{Ivanova2008} may also contribute. These alternative channels may produce different eccentricity distributions and, if a tertiary companion is observed, the inclination between inner and outer orbits may help to distinguish EKL-induced systems from those formed through other mechanisms. }

{Notably, the inclusion of dynamical tides plays a crucial role in the formation of UCXBs. For example, when WD tides are neglected, a comparable fraction of BH–WD systems ($\sim 4.7\%$) can still undergo strong EKL oscillation and lose orbital energy in the simulation. However, instead of gradually shrinking their orbits and forming nearly-circular UCXBs, the majority ($>90\%$) of these systems reach the Roche limit at pericenter ($r_p \lesssim 10^{-4}$~au) while retaining very high eccentricities ($e > 0.9999$, see, e.g., \citet{Fragione_2020} for a similar result). In other words, omitting WD tides can lead to unphysical outcomes, in which many systems behave as “direct plunge-in” mergers or highly eccentric tidal disruption events (TDEs), rather than stable mass transfer UCXBs.
}

Furthermore, previous studies have shown that compact binaries containing a WD can undergo significant tidal heating prior to merger. This process can brighten the WD, potentially trigger runaway fusion in its hydrogen envelope, and lead to a tidal nova \citep{Fuller2012tidalnovae,Fuller2013MNRAS,Vick2017}. In our simulations, we have a similar effect in UCXB progenitor systems: when the orbital eccentricity is significant and the WD has not yet synchronized its spin, the tidal heating rate can reach up to $\sim10^{34}$–$10^{36}$erg~s$^{-1}$. In other words, during the dynamical evolution of triple BH–WD systems, tidal dissipation may heat the hydrogen layer of the WD and ignite it before the orbit evolves into the UCXB phase, which could yield observable signatures. For simplicity, we neglect the effect of tidal heating on the internal structure of the WD and adopt a fixed-temperature WD model when computing the tidal torque. However, we note that heating the WD may reduce the dimensionless tidal torque, thereby further suppressing the overall tidal dissipation after the highly-eccentric formation stage \citep[see, e.g.,][]{Fuller2012tidalnovae,Fuller2013MNRAS}. Conversely, highly eccentric WD binaries may also undergo chaotic tides, where repeated pericenter passages stochastically excite large-amplitude stellar oscillations before significant damping, enhancing tidal heating and orbital decay \citep[e.g., ][]{Vick2017,lau2025}. Future work may improve this model by incorporating thermal feedback and chaotic tides into the WD evolution.

Additionally, our simulations adopt a representative configuration for all BH-WD systems ($10$~M$_{\odot}$ for the BH and $0.6$~M$_{\odot}$ for the WD, with dimensionless tidal torque $\hat{F}(\omega)\sim 150\omega^5$), following \citet{shariat2025triple}. However, in reality, BH–WD binaries span a range of component masses and WD temperatures, which inevitably influences their orbital evolution and the detectability in GW and X-rays. Nonetheless, our results for the triple formation channel remain robust. {In particular, the point mass EKL dynamics are largely insensitive to the details of the mass ratio $q$ as long as it is small, such as in the case of BH and WD progenitors \citep{naoz13,Naoz16,stephan16}.} Also, white dwarf masses in compact binaries are strongly clustered around $0.5-0.7$~M$_{\odot}$\citep[see, e.g.,][]{Toonen_2018nswd, Xuan+21}, and recent Gaia observations confirm that stellar-mass BH in wide binaries typically lie within the few- to tens-of-solar-mass regime (e.g., {\it Gaia} BH1 and BH2 have masses of $9.62 \pm 0.18 M_{\odot}$ and $8.94 \pm 0.34 M_{\odot}$ respectively, see \citet[][]{Chakrabarti2023gaiaBH1,ElBadry23_BH2}). In addition, we also test a wide range of WD temperatures (i.e., dimensionless tidal torque prescriptions, from $\hat{F}(\omega)\sim 10\omega^5$ to $\sim 1000\omega^5$), and find that our simulation results are not significantly affected, with the resultant UCXB formation rate and lifetime varying by less than a factor of two.

{Notably, adopting a realistic mass distribution yields a comparable UCXB fraction. For example, we resample the component masses of BH–WD systems from progenitor distributions of $20<m_1/{\rm M_\odot}<40$, $0.8<m_2/{\rm M_\odot}<8$, and $0.5<m_3/{\rm M_\odot}<8$, motivated by the observed Gaia triple population (see the discussion below Equation~(\ref{eq:formation_rate})). This sample results in a UCXB fraction of $f_{\rm UCXB}\sim 9\%$, close to the fraction obtained from our fiducial model ($f_{\rm UCXB}\sim 6.3\%$). {We also tested extreme cases by fixing $m_1=30~{\rm M_\odot}$ or $m_3=0.1~{\rm M_\odot}$ in all simulated systems, which gives $f_{\rm UCXB}\sim 4.8\%$ and $1.5\%$, respectively.} These tests confirm that the EKL+tides channel can robustly form BH–WD UCXBs, despite uncertainties in the initial mass distribution of Galactic triples, and predict source numbers consistent in order of magnitude with the observed UCXB population in the Milky Way.} 

We note that, the magnitude and prevalence of BH natal kicks remain extremely uncertain. Some theoretical models predict negligible kicks from direct collapse \citep[e.g.,][]{Woosley95,Sukhbold16,Mirabel17}, while observational studies of BH–LMXBs indicate a mixed picture, with some systems receiving little to no kick and others experiencing large ones \citep[e.g.,][see the latter for a recent overview]{Fragos09,Reid14,Mirabel17,Shenar22,Andrews22,Kimball23,Dashwood24,Burdge24,shariat2025triple,nagarajan2025kick}. However, the only confirmed BH-LMXB in a triple, V404 Cygni, formed with a negligible kick \citep[$v_{\rm k}\lsim5~{\rm km~s^{-1}}$][]{burdge2024,shariat2025triple}, providing justification for our baseline assumption.

For the initial population used here, \citet{shariat2025triple} also modeled triples with natal kick magnitudes ($v_{\rm k}$) drawn from a Maxwellian distribution with $\sigma = 265~{\rm kms^{-1}}$ (as for neutron stars) scaled by the BH gravitational mass \citep{Hansen97,Arzoumanian02,Hobbs04}. Their results, with this assumed $v_{\rm k}$ model, show that kicks primarily unbind wide triples but keep the orbital properties of those that survive largely unchanged from the no-kick population. In this framework, the main effect of including kicks is to scale the predicted rates in Equation~(\ref{eq:formation_rate}) by the survival fraction $f_{\rm survive}$. For example, in the extreme high-kick case above, $f_{\rm survive} \approx 2\%$ \citep[][]{shariat2025triple}, whereas a more realistic scenario in which some BHs receive large kicks and others receive almost none \citep[][]{nagarajan2025kick} would yield a fraction closer to $f_{\rm survive} \approx 50\%$. 

Given that the only known BH in a triple formed without a natal kick \citep[V404 Cygni;][]{burdge2024,shariat2025triple}, there is strong empirical support for at least a subset of BHs forming with low or negligible kicks, implying that a substantial population of detached BH–WD inner binaries in triples should exist in the Galactic stellar population.

\section{Conclusions}
\label{sec:conclusion}
Ultracompact X-ray binaries are important multi-messenger sources emitting both X-rays and gravitational wave signals.  Observational evidence has confirmed their existence, making them valuable targets for probing compact binary evolution. In this work, we focus on the evolution of detached BH–WD binaries in the Galactic field, and show that they can naturally form UCXBs through the dynamical perturbation of distant tertiary companions. 

In particular, based on the previous study of massive main-sequence triples, we generate an observationally motivated BH-WD population with tertiary companions (see Section~\ref{subsec:initial} and Figure~\ref{fig:initial} in Appendix~\ref{sec:appendixA}). We then simulated their evolution (see Sections~\ref{subsec:dyntides} and \ref{subsec:masstransfer}), including the eccentric Kozai–Lidov mechanism, general relativistic precession, gravitational wave emission, WD dynamical tides, and WD mass transfer. We find that a significant fraction ($\sim 6.3\%$) of BH-WDs are perturbed by the EKL effect and excited to extreme eccentricities. Such highly eccentric orbits produce strong dynamical tides and gravitational wave radiation at pericenter (Figure~\ref{fig:tides}). Therefore, the BH-WD binary can undergo rapid orbital decay and circularization (Figure~\ref{fig:eg1}), which in turn initiates mass transfer and transforms the system into a UCXB. We further explore the evolution track of these triple-induced UCXBs (see Figure~\ref{fig:population}), and couple the time evolution of their GW signatures with the EM counterpart. Our simulation shows these BH-UCXBs have a formation rate of $\sim 1.5\times 10^{-7}$~yr$^{-1}$ in the Milky Way, with an average GW detectable time of $\sim 4$~Myr and X-ray detectable time $\sim 20-180$~Myr. 

To conclude, the triple formation channel can yield a promising number of BH-WD sources detectable via both GW ($\sim 1$ source in the MW) and X-rays ($\sim 3-27$ sources in the MW, $\sim 1-5$ sources in M31), which offers a unique opportunity for multimessenger studies. We further suggest that electromagnetic observations searching for distant tertiary companions ($a_2\sim 500-15000$~au) in known UCXBs could serve as a direct test of their triple origin. The detection of UCXBs formed via this channel would not only shed light on the dynamical evolution of field triples but also help constrain the physics of WD dynamical tides and contribute a valuable population of LISA verification binaries for future gravitational wave observation.




\acknowledgments
{The authors thank the anonymous referee for their valuable feedback and helpful suggestions.} We also thank Jim Fuller and Linhao Ma for the valuable discussions, and thank the 56th Annual Meeting of the Division on Dynamical Astronomy (DDA) for sparking early conversations about these topics.
ZX acknowledges partial support from the Bhaumik Institute for Theoretical Physics summer fellowship.
ZX and SN acknowledge the partial support from 
NSF-AST Grant No. 2206428, and thank Howard and Astrid Preston for their generous support. C.S. acknowledges support from the Joshua and Beth Friedman Foundation Fund and the Department of Energy Computational Science Graduate Fellowship supported by the U.S. Department
of Energy, Office of Science, Office of Advanced Scientific Computing Research, under Award Number DE-SC0026073. This research was supported by the Munich Institute for Astro-, Particle and BioPhysics (MIAPbP), which is funded by the Deutsche Forschungsgemeinschaft (DFG, German Research Foundation) under Germany´s Excellence Strategy – EXC-2094 – 390783311.

\appendix
\section{Initial Conditions}
\label{sec:appendixA}
\begin{figure*}[htbp]
    \centering
    \includegraphics[width=3.5in]{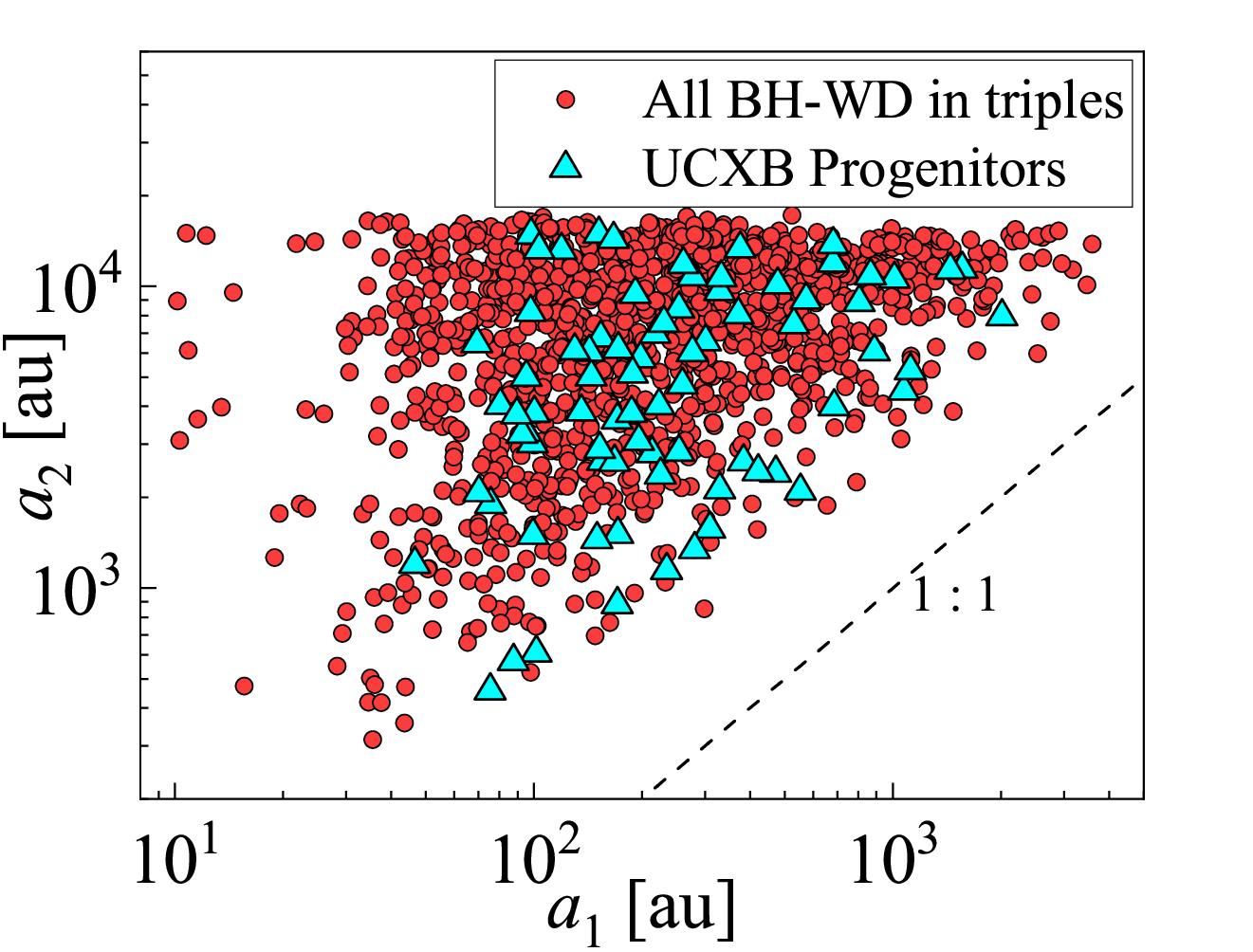}
    \caption{{\bf Parameters of detached BH-WD systems in triples after main sequence evolution.} Here we plot the semimajor axis of inner ($a_1$) and outer orbits ($a_2$) for the BH–WD binaries and their tertiary companions, immediately after the inner binary has evolved into a BH–WD configuration. { Red circles represent all systems, while blue triangles indicate those that later form UCXBs in our simulations.} The dashed line indicates $a_2/a_1=1$.} 
    \label{fig:initial}
\end{figure*}
Here, we show the initial separation of detached BH-WDs versus the outer orbit semi-major axis of their tertiary companions. These systems have survived the main sequence evolution in \citet{shariat2025triple}. Notably, because the inner binaries of these systems have lost mass when becoming a BH-WD system, their inner orbits are, in general, wider than the separation during main-sequence evolution (see Figure~\ref{fig:initial}). Such a wider separation enhances the EKL effect, making the BH-WD more likely to be excited to high eccentricity during the subsequent evolution.

\section{An analytical approximation of WD dynamical tides}
\label{sec:appendixB}
As discussed in Section~\ref{subsec:dyntides}, the tidal energy and angular momentum transfer rate of a BH-WD system can be computed using Equations~(\ref{eq:Jdot}) and (\ref{eq:Edot}). However, for highly eccentric stages during dynamical evolution, the summation in these two equations may include thousands to millions of harmonics, which is computationally expensive. Thus, here we further fit the results of Equations~(\ref{eq:Jdot}) and (\ref{eq:Edot}) using empirical formulas.

In particular, when fixing the system mass, $\dot{E}_{\rm tide}$ and $\dot{J}_{\rm tide}$ can be expressed as functions of the orbital eccentricity $e$, pericenter distance $r_p$, and WD spin frequency $\Omega_s$. For simplicity, we start by fitting the zero-spin case. Notably, $\dot{E}_{\rm tide}(\Omega_s=0)$ and $\dot{J}_{\rm tide}(\Omega_s=0)$ have a straightforward power law relation to the pericenter distance $r_p$ (see, e.g., Equations~(\ref{eq:Jdot}) - (\ref{eq:Edot}) and Figure~\ref{fig:tides}):
\begin{equation}
\left\{\begin{array}{l}\dot{E}_{\text {tide }}\left(\Omega_s=0\right)\propto T_0\cdot\Omega\cdot\hat{F}(\Omega)\propto a^{-15}\sim r_p^{-15} \\
\dot{J}_{\text {tide }}\left(\Omega_s=0\right)\propto T_0\cdot\hat{F}(\Omega)\propto a^{-13.5}\sim r_p^{-13.5} 
\end{array}\right. \ .
\label{eq:rppowerlaw}
\end{equation}
This relation is understood because the WD's dimensionless tidal torque has an approximated power law relation to the orbital frequency \citep[$\hat{F}(\omega)\sim 150\omega^5$ for a 5000K WD in $G=M=R=1$ unit, see, e.g.,][]{Vick2017}. Furthermore, changing $r_p$ (or effectively the orbital separation $a$) simultaneously changes the terms of $\omega\sim N\Omega$ in Equations~(\ref{eq:Jdot}) and (\ref{eq:Edot}). 

Equation~(\ref{eq:rppowerlaw}) highlights the strong dependence of dynamical tides on the pericenter distance, which also indicates that $r_p$ most strongly determines the order-of-magnitude of tidal effects. Furthermore, given the same $r_p$, orbits with different eccentricities can have different tidal torque magnitudes, which is reflected in the Hansen coefficients and sum of harmonics in Equations~(\ref{eq:Jdot}) and (\ref{eq:Edot}). To estimate this effect, we numerically computed the magnitude of $\dot{E}_{\rm tide}$, $\dot{J}_{\rm tide}$, and fit them as functions of eccentricity e. Empirically, we find that, instead of fitting the exact values of $\dot{E}_{\rm tide}$ and $\dot{J}_{\rm tide}$, the results are better approximated when fitting the ratio between tidal and GW energy (angular momentum) transfer rates \citep[see, e.g., equations 5.4 and 5.5 in][]{Peters64}:\footnote{Note that here we only fit the empirical equations for $0.2<e<1$, as the nearly circular system has few number of harmonics and the analytical appriximation breaks. Furthermore, here we specifically fit the coefficient for the $10-0.6$~M$_{\odot}$ BH-WD system, with $R_2=5.3\times10^{-5}$~au and dimensionless tidal torque $\hat{F}(\omega)=150\omega^5$. However, it is straightforward to rescale the result for other configurations. For example, when fixing $a$ and $e$ and changing $m_1$, $\dot{E}$ scales as $m_1^2(m_1+m_2)^3$ and $\dot{J}$ scales as $m_1^2(m_1+m_2)^{5/2}$, since $m_1$ affects both $T_0$ and orbital frequency. }
\begin{equation}
    \frac{\dot{E}_{\rm tide}(\Omega_s =0)}{\dot{E}_{\rm GW}}  \sim (92.7+1204e+3124e^2)\times \left(\frac{r_p}{10^{-3}~{\rm au}}\right)^{-10}
    \label{eq:Edotana}
\end{equation}
\begin{equation}
    \frac{\dot{J}_{\rm tide}(\Omega_s =0)}{\dot{J}_{\rm GW}}  \sim (105+691e+708e^2)\times \left(\frac{r_p}{10^{-3}~{\rm au}}\right)^{-10}
    \label{eq:Jdotana}
\end{equation}

\begin{figure*}[htbp]
    \centering
    \includegraphics[width=3.5in]{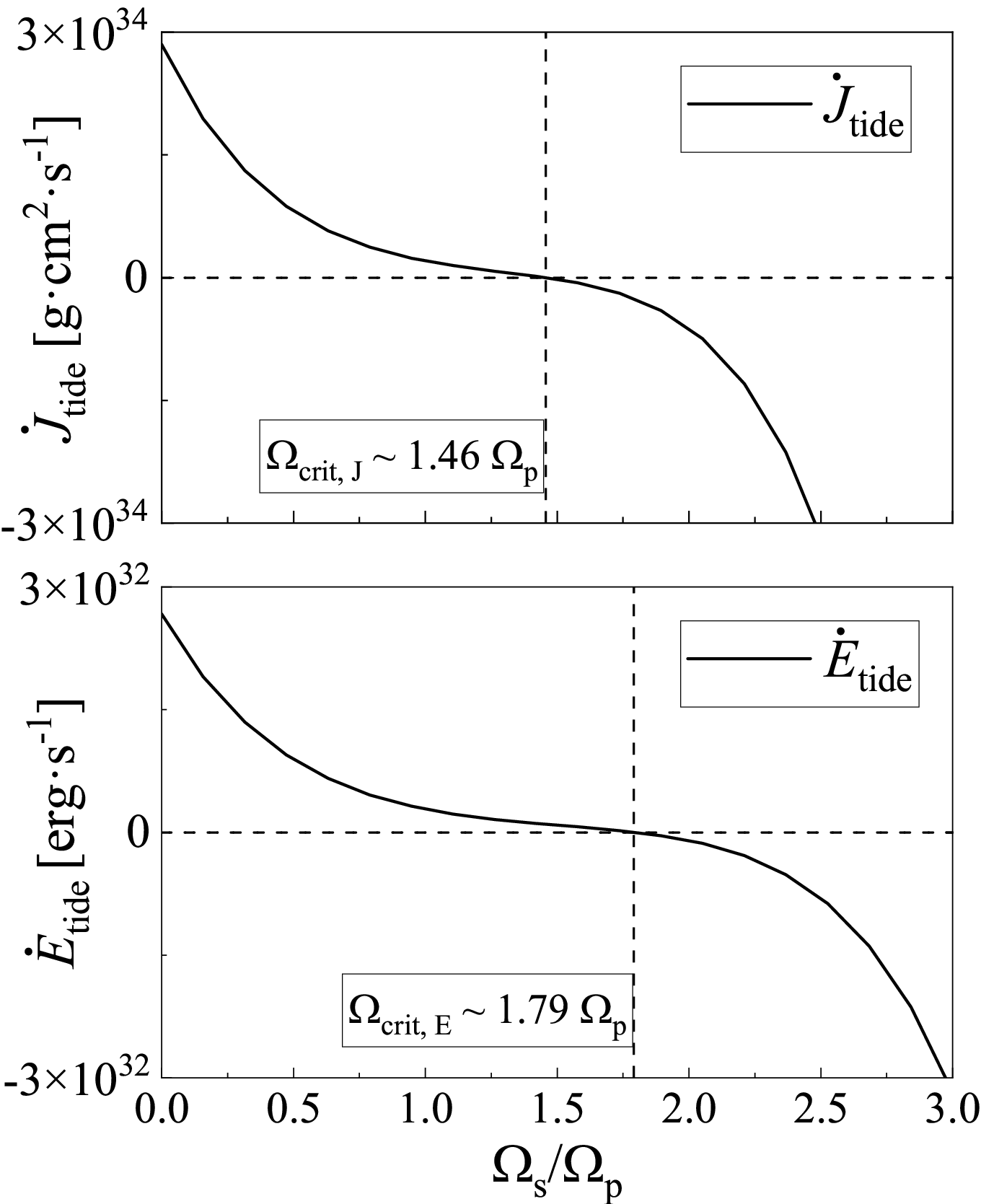}
    \caption{{\bf Comparison between tidal angular momentum and energy transfer rate, $\dot{J}_{\rm tide}$ and $\dot{E}_{\rm tide}$, as functions of the WD spin.} Here we consider a $10-0.6$~M$_{\odot}$ BH-WD system with $e=0.9,\, r_p =3\times10^{-3}$~au, $\hat{F}(\omega)=150\omega^5$, and plot $\dot{J}_{\rm tide}$ and $\dot{E}_{\rm tide}$ as functions of the ratio between the WD spin angular frequency, $\Omega_s$, and orbital pericenter frequency, $\Omega_p$.} 
    \label{fig:omegacrit}
\end{figure*}

Next, we consider the general case by releasing the assumption of $\Omega_s=0$. Notably, the evolution of $\Omega_s$ significantly affects $\dot{E}_{\rm tide}$ and $\dot{J}_{\rm tide}$. For example, Figure~\ref{fig:omegacrit} plots $\dot{J}_{\rm tide}$ and $\dot{E}_{\rm tide}$ as functions of the ratio between the WD spin angular frequency $\Omega_s$ and orbital pericenter frequency $\Omega_p$ (computed strictly using Equations~(\ref{eq:Jdot}) and (\ref{eq:Edot}) for a $e=0.9$ system). As shown in the figure, starting from the zero-spin case, the magnitude of dynamical tides is suppressed as a WD spins up. Consequently, the WD can spin up to the position of $\Omega_{\rm crit, J}$, where the tidal torque stops transferring angular momentum ($\dot{J}=0$) and the spin pseudo-synchronizes with the orbit. This phenomenon also applies to the energy transfer rate, with $\Omega_{\rm crit, E}$ representing the critical spin when there is no net tidal energy loss from the orbit. However, we note that $\Omega_{\rm crit, J}$ is smaller than $\Omega_{\rm crit, E}$ when $e>0$. In other words, once an eccentric system get tidally locked at $\Omega_s=\Omega_{\rm crit, J}$, it still has a residual energy transfer rate (at $\Omega_s<\Omega_{\rm crit, E}$, see, e.g., the position of $\Omega_s=1.46\Omega_p$ in the figure). This is because eccentricity always excites multiple harmonics of dynamical tides and causes tidal heating, see the discussion in Section~\ref{subsec:dyntides}.

We characterize the effect of $\Omega_s$ on $\dot{E}_{\rm tide}$ and $\dot{J}_{\rm tide}$ using an extra suppression factor $K$ ($K_E$ for the energy, $K_J$ for the angular momentum, respectively). In particular, when $\Omega_s<\Omega_{\rm s,\,crit}$, we have:
\begin{equation}
\left\{\begin{array}{l}\dot{E}_{\text {tide }}(\Omega_s)=\dot{E}_{\text {tide }}(\Omega_s=0) \cdot [K_E(\Omega_s, e)-K_E(\Omega_{\rm s,\,Ecrit}, e)]/[K_E(0, e)-K_E(\Omega_{\rm s,\,Ecrit}, e)]\\
\dot{J}_{\text {tide }}(\Omega_s)=\dot{J}_{\text {tide }}(\Omega_s=0) \cdot [K_J(\Omega_s, e)-K_J(\Omega_{\rm s,\,Jcrit}, e)] /[K_J(0, e)-K_J(\Omega_{\rm s,\,Jcrit}, e)]
\label{eq:app1}
\end{array}\right.
\end{equation}
where $K_E,\,K_J$ are the suppression factors. They can be approximated using the following relations (fitted for $0.2<e<1$):
\begin{equation}
\left\{\begin{array}{l}K_E\left(\Omega_s, e\right) \sim \exp\{[-5.23+9.06e-10.1e^2+4.34e^3]\cdot(\Omega_{\rm s}/\Omega_p)\}\\
K_J\left(\Omega_s, e\right) \sim \exp\{[-5.70+9.82e-11.0e^2+4.79e^3]\cdot(\Omega_{\rm s}/\Omega_p)\}
\end{array}\right.
\label{eq:app2}
\end{equation}
and $\Omega_{\rm s,\,Ecrit}$, $\Omega_{\rm s,\,Jcrit}$ represent the critical WD spin rate at which the energy flux and the angular momentum flux equal zero, respectively, which can be estimated using:
\begin{equation}
\left\{\begin{array}{l}\Omega_{\rm s,\,Ecrit}/\Omega_p\sim 1+1.10e-1.75e^2+5.28e^3-6.23e^4+2.47e^5\\
\Omega_{\rm s,\,Jcrit}/\Omega_p\sim 1+0.886e-3.95e^2+10.9e^3-11.6e^4+4.3e^5
\end{array}\right.
\label{eq:app3}
\end{equation}

We note that Equations~(\ref{eq:app1}) - (\ref{eq:app3}) is only valid for $\Omega_s<\Omega_{\rm crit}$. On the other hand, the WD spin in principle can exceed $\Omega_{\rm s,\,Jcrit}$ ($\Omega_{\rm s,\,Ecrit}$), thus transferring angular momentum (energy) back to the orbit. For completeness, we also fit the value of $\dot{E}$ and $\dot{J}$ at $\Omega_s>\Omega_{\rm s,\,crit}$. 
\begin{equation}
\left\{\begin{array}{l}\dot{E}_{\text {tide }}(\Omega_s)=\dot{E}_{\text {tide }}(\Omega_s=0) \cdot [K_E'(\Omega_s, e)-K_E'(\Omega_{\rm s,\,Ecrit}, e)]\\
\dot{J}_{\text {tide }}(\Omega_s)=\dot{J}_{\text {tide }}(\Omega_s=0) \cdot [K_J'(\Omega_s, e)-K_J'(\Omega_{\rm s,\,Jcrit}, e)]
\end{array}\right.
\end{equation}
where $K_E',\,K_J'$ are the factor controlling the growth of tidal effect when $\Omega_s>\Omega_{\rm s,\,Ecrit},\,\Omega_s>\Omega_{\rm s,\,Ecrit}$, respectively:
\begin{equation}
\left\{\begin{array}{l}K_E'\left(\Omega_s, e\right) \sim -(0.652-2.10e+2.46e^2-1.00e^3)[(\Omega_s/\Omega_p)-(0.620+0.889e-1.27e^2+0.627e^3)]^5\\
K_J'\left(\Omega_s, e\right) \sim -(1.48-5.38e+7.22e^2-3.33e^3)[(\Omega_s/\Omega_p)-(1.07-2.44e+6.02e^2-6.37e^3+2.55e^4-0.112e^5)]^5
\end{array}\right.
\end{equation}


\section{X-ray Detectable time of UCXBs}\label{sec:appendixC}
\begin{figure*}[htbp]
    \centering
    \includegraphics[width=3.5in]{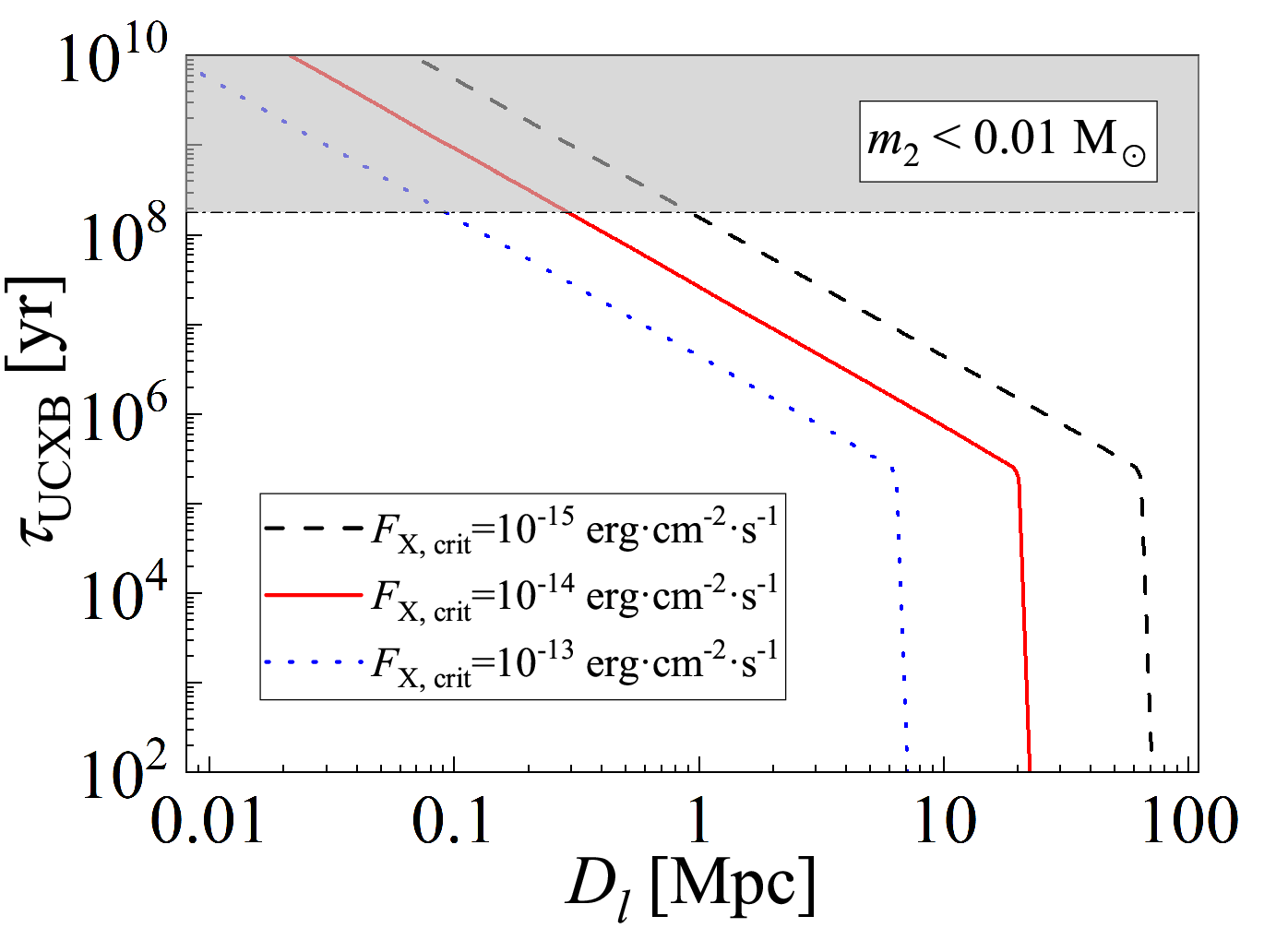}
    \caption{{\bf The X-ray detectable time of a representative UCXB source, as a function of its distance.} Here we consider a mass-transferring BH-WD system with initial mass $m_1=10$~$\rm M_{\odot}$, $m_2=0.6$~$\rm M_{\odot}$, and computed its total detectable lifetime in X-ray, assuming a detector sensitivity of $F_{\rm X, crit}=10^{-15},\, 10^{-14},\,10^{-13}$~erg~cm$^{-2}$~s$^{-1}$, respectively (from left to right).} 
    \label{fig:dmax_xray}
\end{figure*}

Here, we adopt the simulated X-ray luminosity evolution of BH-UCXB systems (see, e.g., Figure~\ref{fig:eg1}), and evaluate their detectable lifetime. In particular, we show the maximum detectable time of a $10-0.6$~M$_{\odot}$ UCXB as a function of its distance to the observer, assuming the X-ray detector has a limiting flux of $F_{\rm X, crit}=10^{-15},\, 10^{-14},\,10^{-13}$~erg~cm$^{-2}$~s$^{-1}$, respectively. This range of limiting fluxes is representative of modern present-day X-ray observatories, such as Chandra, XMM-Newton, or eROSITA.The grey-shaded region in Figure~\ref{fig:dmax_xray} represents where the WD donor has a mass lower than $0.01$~M$_{\odot}$. Such a light mass may cause instability and stop the mass transfer evolution (see Section~\ref{subsec:masstransfer}), thus we exclude those systems in our work, for conservative purposes. The figure demonstrates that even out to $D_{\rm l}\gsim10-100$~Mpc, these UCXBs can be observed as transient X-ray sources among all-sky surveys.





\bibliography{bibbase}

@ARTICLE{Michaely+20,
       author = {{Michaely}, Erez and {Perets}, Hagai B.},
        title = "{High rate of gravitational waves mergers from flyby perturbations of wide black hole triples in the field}",
      journal = {\mnras},
         year = 2020,
        month = nov,
       volume = {498},
       number = {4},
        pages = {4924-4935},
          doi = {10.1093/mnras/staa2720},
archivePrefix = {arXiv},
       eprint = {2008.01094},
 primaryClass = {astro-ph.HE},
       adsurl = {https://ui.adsabs.harvard.edu/abs/2020MNRAS.498.4924M}
}

@ARTICLE{Wen03,
       author = {{Wen}, Linqing},
        title = "{On the Eccentricity Distribution of Coalescing Black Hole Binaries Driven by the Kozai Mechanism in Globular Clusters}",
      journal = {\apj},
         year = 2003,
        month = nov,
       volume = {598},
       number = {1},
        pages = {419-430},
          doi = {10.1086/378794},
archivePrefix = {arXiv},
       eprint = {astro-ph/0211492},
 primaryClass = {astro-ph},
       adsurl = {https://ui.adsabs.harvard.edu/abs/2003ApJ...598..419W}
}

@ARTICLE{amaro17,
   author = {{Amaro-Seoane}, P. and {Audley}, H. and {Babak}, S. and {Baker}, J. and 
	{Barausse}, E. and {Bender}, P. and {Berti}, E. and {Binetruy}, P. and 
	{Born}, M. and {Bortoluzzi}, D. et al.
	},
    title = "{Laser Interferometer Space Antenna}",
  journal = {ArXiv e-prints},
archivePrefix = "arXiv",
   eprint = {1702.00786},
 primaryClass = "astro-ph.IM",
     year = 2017,
    month = feb,
   adsurl = {http://adsabs.harvard.edu/abs/2017arXiv170200786A}
}

@ARTICLE{amaro+22,
       author = {{Amaro-Seoane}, Pau and others},
        title = "{Astrophysics with the Laser Interferometer Space Antenna}",
      journal = {arXiv e-prints},
         year = 2022,
        month = mar,
          eid = {arXiv:2203.06016},
        pages = {arXiv:2203.06016},
archivePrefix = {arXiv},
       eprint = {2203.06016},
 primaryClass = {gr-qc},
       adsurl = {https://ui.adsabs.harvard.edu/abs/2022arXiv220306016A}
}

@ARTICLE{Robson+19,
       author = {{Robson}, Travis and {Cornish}, Neil J. and {Liu}, Chang},
        title = "{The construction and use of LISA sensitivity curves}",
      journal = {Classical and Quantum Gravity},
         year = 2019,
        month = may,
       volume = {36},
       number = {10},
          eid = {105011},
        pages = {105011},
          doi = {10.1088/1361-6382/ab1101},
archivePrefix = {arXiv},
       eprint = {1803.01944},
 primaryClass = {astro-ph.HE},
       adsurl = {https://ui.adsabs.harvard.edu/abs/2019CQGra..36j5011R}
}

@ARTICLE{Michaely+22,
       author = {{Michaely}, Erez and {Naoz}, Smadar},
        title = "{Ultrawide Black Hole-Neutron Star Binaries as a Possible Source for Gravitational Waves and Short Gamma-Ray Bursts}",
      journal = {\apj},
         year = 2022,
        month = sep,
       volume = {936},
       number = {2},
          eid = {184},
        pages = {184},
          doi = {10.3847/1538-4357/ac8a92},
archivePrefix = {arXiv},
       eprint = {2205.15040},
 primaryClass = {astro-ph.HE},
       adsurl = {https://ui.adsabs.harvard.edu/abs/2022ApJ...936..184M}
}

@ARTICLE{Naoz+13,
       author = {{Naoz}, Smadar and {Farr}, Will M. and {Lithwick}, Yoram and {Rasio}, Frederic A. and {Teyssandier}, Jean},
        title = "{Secular dynamics in hierarchical three-body systems}",
      journal = {\mnras},
         year = 2013,
        month = may,
       volume = {431},
       number = {3},
        pages = {2155-2171},
          doi = {10.1093/mnras/stt302},
archivePrefix = {arXiv},
       eprint = {1107.2414},
 primaryClass = {astro-ph.EP},
       adsurl = {https://ui.adsabs.harvard.edu/abs/2013MNRAS.431.2155N}
}

@ARTICLE{Wang+21,
       author = {{Wang}, Huiyi and {Stephan}, Alexander P. and {Naoz}, Smadar and {Hoang}, Bao-Minh and {Breivik}, Katelyn},
        title = "{Gravitational-wave Signatures from Compact Object Binaries in the Galactic Center}",
      journal = {\apj},
         year = 2021,
        month = aug,
       volume = {917},
       number = {2},
          eid = {76},
        pages = {76},
          doi = {10.3847/1538-4357/ac088d},
archivePrefix = {arXiv},
       eprint = {2010.15841},
 primaryClass = {astro-ph.HE},
       adsurl = {https://ui.adsabs.harvard.edu/abs/2021ApJ...917...76W}
}

@ARTICLE{Hoang+18,
       author = {{Hoang}, Bao-Minh and {Naoz}, Smadar and {Kocsis}, Bence and {Rasio}, Frederic A. and {Dosopoulou}, Fani},
        title = "{Black Hole Mergers in Galactic Nuclei Induced by the Eccentric Kozai-Lidov Effect}",
      journal = {\apj},
         year = 2018,
        month = apr,
       volume = {856},
       number = {2},
          eid = {140},
        pages = {140},
          doi = {10.3847/1538-4357/aaafce},
archivePrefix = {arXiv},
       eprint = {1706.09896},
 primaryClass = {astro-ph.HE},
       adsurl = {https://ui.adsabs.harvard.edu/abs/2018ApJ...856..140H}
}

@article{Deme+20,
	doi = {10.3847/1538-4357/abafa3},
	url = {https://doi.org/10.3847%2F1538-4357%2Fabafa3},
	year = 2020,
	month = {sep},
	publisher = {American Astronomical Society},
	volume = {901},
	number = {2},
	pages = {125},
	author = {Barnab{\'{a}}s Deme and Bao-Minh Hoang and Smadar Naoz and Bence Kocsis},
	title = {Detecting Kozai{\textendash}Lidov Imprints on the Gravitational Waves of Intermediate-mass Black Holes in Galactic Nuclei},
	journal = {The Astrophysical Journal}
}

@ARTICLE{Moe+17,
       author = {{Moe}, Maxwell and {Di Stefano}, Rosanne},
        title = "{Mind Your Ps and Qs: The Interrelation between Period (P) and Mass-ratio (Q) Distributions of Binary Stars}",
      journal = {\apjs},
         year = 2017,
        month = jun,
       volume = {230},
       number = {2},
          eid = {15},
        pages = {15},
          doi = {10.3847/1538-4365/aa6fb6},
archivePrefix = {arXiv},
       eprint = {1606.05347},
 primaryClass = {astro-ph.SR},
       adsurl = {https://ui.adsabs.harvard.edu/abs/2017ApJS..230...15M}
}

@ARTICLE{Sana+12,
       author = {{Sana}, H. and {de Mink}, S.~E. and {de Koter}, A. and {Langer}, N. and {Evans}, C.~J. and {Gieles}, M. and {Gosset}, E. and {Izzard}, R.~G. and {Le Bouquin}, J. -B. and {Schneider}, F.~R.~N.},
        title = "{Binary Interaction Dominates the Evolution of Massive Stars}",
      journal = {Science},
         year = 2012,
        month = jul,
       volume = {337},
       number = {6093},
        pages = {444},
          doi = {10.1126/science.1223344},
archivePrefix = {arXiv},
       eprint = {1207.6397},
 primaryClass = {astro-ph.SR},
       adsurl = {https://ui.adsabs.harvard.edu/abs/2012Sci...337..444S}
}

@INPROCEEDINGS{Tokovinin+08,
       author = {{Tokovinin}, A. and {Thomas}, S. and {Sterzik}, M. and {Udry}, S.},
        title = "{Tertiary Companions to Close Spectroscopic Binaries}",
    booktitle = {Multiple Stars Across the H-R Diagram},
         year = 2008,
       editor = {{Hubrig}, Swetlana and {Petr-Gotzens}, Monika and {Tokovinin}, Andrei},
        month = jan,
        pages = {129},
          doi = {10.1007/978-3-540-74745-1_19},
       adsurl = {https://ui.adsabs.harvard.edu/abs/2008msah.conf..129T}
}

@ARTICLE{Hamers+18,
       author = {{Hamers}, Adrian S.},
        title = "{Secular dynamics of hierarchical multiple systems composed of nested binaries, with an arbitrary number of bodies and arbitrary hierarchical structure - II. External perturbations: flybys and supernovae}",
      journal = {\mnras},
         year = 2018,
        month = may,
       volume = {476},
       number = {3},
        pages = {4139-4161},
          doi = {10.1093/mnras/sty428},
archivePrefix = {arXiv},
       eprint = {1802.05716},
 primaryClass = {astro-ph.SR},
       adsurl = {https://ui.adsabs.harvard.edu/abs/2018MNRAS.476.4139H}
}

@ARTICLE{Stephan+19,
       author = {{Stephan}, Alexander P. and {Naoz}, Smadar and {Ghez}, Andrea M. and
         {Morris}, Mark R. and {Ciurlo}, Anna and {Do}, Tuan and
         {Breivik}, Katelyn and {Coughlin}, Scott and {Rodriguez}, Carl L.},
        title = "{The Fate of Binaries in the Galactic Center: The Mundane and the Exotic}",
      journal = {\apj},
         year = 2019,
        month = jun,
       volume = {878},
       number = {1},
          eid = {58},
        pages = {58},
          doi = {10.3847/1538-4357/ab1e4d},
archivePrefix = {arXiv},
       eprint = {1903.00010},
 primaryClass = {astro-ph.SR},
       adsurl = {https://ui.adsabs.harvard.edu/abs/2019ApJ...878...58S}
}

@ARTICLE{2016PhRvD..93b4003K,
       author = {{Klein}, Antoine and {Barausse}, Enrico and {Sesana}, Alberto and
         {Petiteau}, Antoine and {Berti}, Emanuele and {Babak}, Stanislav and
         {Gair}, Jonathan and {Aoudia}, Sofiane and {Hinder}, Ian and
         {Ohme}, Frank and {Wardell}, Barry},
        title = "{Science with the space-based interferometer eLISA: Supermassive black hole binaries}",
      journal = {\prd},
         year = 2016,
        month = jan,
       volume = {93},
       number = {2},
          eid = {024003},
        pages = {024003},
          doi = {10.1103/PhysRevD.93.024003},
archivePrefix = {arXiv},
       eprint = {1511.05581},
 primaryClass = {gr-qc},
       adsurl = {https://ui.adsabs.harvard.edu/abs/2016PhRvD..93b4003K}
}

@ARTICLE{Xuan+21,
       author = {{Xuan}, Zeyuan and {Peng}, Peng and {Chen}, Xian},
        title = "{Degeneracy between mass and peculiar acceleration for the double white dwarfs in the LISA band}",
      journal = {\mnras},
         year = 2021,
        month = apr,
       volume = {502},
       number = {3},
        pages = {4199-4209},
          doi = {10.1093/mnras/stab331},
archivePrefix = {arXiv},
       eprint = {2012.00049},
 primaryClass = {astro-ph.HE},
       adsurl = {https://ui.adsabs.harvard.edu/abs/2021MNRAS.502.4199X}
}

@ARTICLE{corral16,
   author = {{Corral-Santana}, J.~M. and {Casares}, J. and {Mu{\~n}oz-Darias}, T. and
        {Bauer}, F.~E. and {Mart{\'{\i}}nez-Pais}, I.~G. and {Russell}, D.~M.
        },
    title = "{BlackCAT: A catalogue of stellar-mass black holes in X-ray transients}",
  journal = {\aap},
archivePrefix = "arXiv",
   eprint = {1510.08869},
 primaryClass = "astro-ph.HE",
     year = 2016,
    month = mar,
   volume = 587,
      eid = {A61},
    pages = {A61},
      doi = {10.1051/0004-6361/201527130},
   adsurl = {http://adsabs.harvard.edu/abs/2016A%26A...587A..61C}
}

@ARTICLE{kozai62,
   author = {{Kozai}, Y.},
    title = "{Secular perturbations of asteroids with high inclination and eccentricity}",
  journal = {\aj},
     year = 1962,
    month = nov,
   volume = 67,
    pages = {591},
      doi = {10.1086/108790},
   adsurl = {http://adsabs.harvard.edu/abs/1962AJ.....67..591K}
}

@ARTICLE{lidov62,
   author = {{Lidov}, M.~L.},
    title = "{The evolution of orbits of artificial satellites of planets under the action of gravitational perturbations of external bodies}",
  journal = {Planetary and Space Science},
     year = 1962,
    month = oct,
   volume = 9,
    pages = {719-759},
      doi = {10.1016/0032-0633(62)90129-0},
   adsurl = {http://adsabs.harvard.edu/abs/1962P%26SS....9..719L}
}

@ARTICLE{mardling01,
   author = {{Mardling}, R.~A. and {Aarseth}, S.~J.},
    title = "{Tidal interactions in star cluster simulations}",
  journal = {\mnras},
     year = 2001,
    month = mar,
   volume = 321,
    pages = {398-420},
      doi = {10.1046/j.1365-8711.2001.03974.x},
   adsurl = {http://adsabs.harvard.edu/abs/2001MNRAS.321..398M}
}

@ARTICLE{naoz13,
   author = {{Naoz}, S. and {Kocsis}, B. and {Loeb}, A. and {Yunes}, N.},
    title = "{Resonant Post-Newtonian Eccentricity Excitation in Hierarchical Three-body Systems}",
  journal = {\apj},
archivePrefix = "arXiv",
   eprint = {1206.4316},
 primaryClass = "astro-ph.SR",
     year = 2013,
    month = aug,
   volume = 773,
      eid = {187},
    pages = {187},
      doi = {10.1088/0004-637X/773/2/187},
   adsurl = {http://adsabs.harvard.edu/abs/2013ApJ...773..187N}
}

@ARTICLE{stephan16,
   author = {{Stephan}, A.~P. and {Naoz}, S. and {Ghez}, A.~M. and {Witzel}, G. and
	{Sitarski}, B.~N. and {Do}, T. and {Kocsis}, B.},
    title = "{Merging binaries in the Galactic Center: the eccentric Kozai-Lidov mechanism with stellar evolution}",
  journal = {\mnras},
archivePrefix = "arXiv",
   eprint = {1603.02709},
 primaryClass = "astro-ph.SR",
     year = 2016,
    month = aug,
   volume = 460,
    pages = {3494-3504},
      doi = {10.1093/mnras/stw1220},
   adsurl = {http://adsabs.harvard.edu/abs/2016MNRAS.460.3494S}
}

@ARTICLE{luo16,
   author = {{Luo}, J. and {Chen}, L.-S. and {Duan}, H.-Z. and {Gong}, Y.-G. and
	{Hu}, S. and {Ji}, J. and {Liu}, Q. and {Mei}, J. and {Milyukov}, V. and
	{Sazhin}, M. and {Shao}, C.-G. and {Toth}, V.~T. and {Tu}, H.-B. and
	{Wang}, Y. and {Wang}, Y. and {Yeh}, H.-C. and {Zhan}, M.-S. and
	{Zhang}, Y. and {Zharov}, V. and {Zhou}, Z.-B.},
    title = "{TianQin: a space-borne gravitational wave detector}",
  journal = {CQG},
archivePrefix = "arXiv",
   eprint = {1512.02076},
 primaryClass = "astro-ph.IM",
     year = 2016,
    month = feb,
   volume = 33,
   number = 3,
      eid = {035010},
    pages = {035010},
      doi = {10.1088/0264-9381/33/3/035010},
   adsurl = {http://adsabs.harvard.edu/abs/2016CQGra..33c5010L}
}

@ARTICLE{Peters64,
    author = {{Peters}, P. C.},
    title = "{Gravitational Radiation and the Motion of Two Point Masses}",
    journal = {Physical Review },
    year = 1964,
    month = nov,
    volume = 136,
    pages = {1224--1232},
    opturl = {http://adsabs.harvard.edu/cgi-bin/nph-bib_query?bibcode=1964PhRv..136.1224P&db_key=PHY}
}

@Article{Ivanov2005,
  author        = {{Ivanov}, P.~B. and {Polnarev}, A.~G. and {Saha}, P.},
  title         = {{The tidal disruption rate in dense galactic cusps containing a supermassive binary black hole}},
  journal       = {\mnras},
  year          = {2005},
  volume        = {358},
  pages         = {1361-1378},
  month         = apr,
  __markedentry = {[xian:]},
  adsnote       = {Provided by the SAO/NASA Astrophysics Data System},
  adsurl        = {http://adsabs.harvard.edu/abs/2005MNRAS.358.1361I},
  doi           = {10.1111/j.1365-2966.2005.08843.x},
  eprint        = {astro-ph/0410610},
  keywords      = {black hole physics, galaxies: nuclei},
}

@ARTICLE{Naoz16,
       author = {{Naoz}, Smadar},
        title = "{The Eccentric Kozai-Lidov Effect and Its Applications}",
      journal = {\araa},
         year = 2016,
        month = sep,
       volume = {54},
        pages = {441-489},
          doi = {10.1146/annurev-astro-081915-023315},
archivePrefix = {arXiv},
       eprint = {1601.07175},
 primaryClass = {astro-ph.EP},
       adsurl = {https://ui.adsabs.harvard.edu/abs/2016ARA&A..54..441N}
}

@ARTICLE{Bub+20,
       author = {{Bub}, Mathew W. and {Petrovich}, Cristobal},
        title = "{Compact-object Mergers in the Galactic Center: Evolution in Triaxial Clusters}",
      journal = {\apj},
         year = 2020,
        month = may,
       volume = {894},
       number = {1},
          eid = {15},
        pages = {15},
          doi = {10.3847/1538-4357/ab8461},
archivePrefix = {arXiv},
       eprint = {1910.02079},
 primaryClass = {astro-ph.HE},
       adsurl = {https://ui.adsabs.harvard.edu/abs/2020ApJ...894...15B}
}

@ARTICLE{Zwick+20,
       author = {{Zwick}, Lorenz and {Capelo}, Pedro R. and {Bortolas}, Elisa and {Mayer}, Lucio and {Amaro-Seoane}, Pau},
        title = "{Improved gravitational radiation time-scales: significance for LISA and LIGO-Virgo sources}",
      journal = {\mnras},
         year = 2020,
        month = jun,
       volume = {495},
       number = {2},
        pages = {2321-2331},
          doi = {10.1093/mnras/staa1314},
archivePrefix = {arXiv},
       eprint = {1911.06024},
 primaryClass = {astro-ph.GA},
       adsurl = {https://ui.adsabs.harvard.edu/abs/2020MNRAS.495.2321Z}
}

@ARTICLE{Xuan+23b,
       author = {{Xuan}, Zeyuan and {Naoz}, Smadar and {Kocsis}, Bence and {Michaely}, Erez},
        title = "{Detecting Gravitational Wave Bursts From Stellar-Mass Binaries in the Milli-hertz Band}",
      journal = {arXiv e-prints},
         year = 2023,
        month = sep,
          eid = {arXiv:2310.00042},
        pages = {arXiv:2310.00042},
          doi = {10.48550/arXiv.2310.00042},
archivePrefix = {arXiv},
       eprint = {2310.00042},
 primaryClass = {astro-ph.HE},
       adsurl = {https://ui.adsabs.harvard.edu/abs/2023arXiv231000042X}
}

@ARTICLE{Kroupa2001,
       author = {{Kroupa}, Pavel},
        title = "{On the variation of the initial mass function}",
      journal = {\mnras},
         year = 2001,
        month = apr,
       volume = {322},
       number = {2},
        pages = {231-246},
          doi = {10.1046/j.1365-8711.2001.04022.x},
archivePrefix = {arXiv},
       eprint = {astro-ph/0009005},
 primaryClass = {astro-ph},
       adsurl = {https://ui.adsabs.harvard.edu/abs/2001MNRAS.322..231K}
}

@ARTICLE{Xuan23acc,
       author = {{Xuan}, Zeyuan and {Naoz}, Smadar and {Chen}, Xian},
        title = "{Detecting accelerating eccentric binaries in the LISA band}",
      journal = {\prd},
         year = 2023,
        month = feb,
       volume = {107},
       number = {4},
          eid = {043009},
        pages = {043009},
          doi = {10.1103/PhysRevD.107.043009},
archivePrefix = {arXiv},
       eprint = {2210.03129},
 primaryClass = {astro-ph.HE},
       adsurl = {https://ui.adsabs.harvard.edu/abs/2023PhRvD.107d3009X}
}

@article{Xuan24bkg,
  title = {Stochastic gravitational wave background from highly-eccentric stellar-mass binaries in the millihertz band},
  author = {{Xuan}, Zeyuan and Naoz, Smadar and Kocsis, Bence and Michaely, Erez},
  journal = {Phys. Rev. D},
  volume = {110},
  issue = {2},
  pages = {023020},
  numpages = {15},
  year = {2024},
  month = {Jul},
  publisher = {American Physical Society},
  doi = {10.1103/PhysRevD.110.023020},
  url = {https://link.aps.org/doi/10.1103/PhysRevD.110.023020}
}

@ARTICLE{Xuan24parameter,
       author = {{Xuan}, Zeyuan and {Naoz}, Smadar and {Li}, Alvin K.~Y. and {Kocsis}, Bence and {Petigura}, Erik and {Knee}, Alan M. and {McIver}, Jess and {Kremer}, Kyle and {Farr}, Will M.},
        title = "{Extracting Astrophysical Information of Highly-Eccentric Binaries in the Millihertz Gravitational Wave Band}",
      journal = {arXiv e-prints},
         year = 2024,
        month = sep,
          eid = {arXiv:2409.15413},
        pages = {arXiv:2409.15413},
          doi = {10.48550/arXiv.2409.15413},
archivePrefix = {arXiv},
       eprint = {2409.15413},
 primaryClass = {astro-ph.HE},
       adsurl = {https://ui.adsabs.harvard.edu/abs/2024arXiv240915413X}
}

@ARTICLE{Vick2017,
       author = {{Vick}, Michelle and {Lai}, Dong and {Fuller}, Jim},
        title = "{Tidal dissipation and evolution of white dwarfs around massive black holes: an eccentric path to tidal disruption}",
      journal = {\mnras},
         year = 2017,
        month = jun,
       volume = {468},
       number = {2},
        pages = {2296-2310},
          doi = {10.1093/mnras/stx539},
       adsurl = {https://ui.adsabs.harvard.edu/abs/2017MNRAS.468.2296V}
}

@ARTICLE{Su2022,
       author = {{Su}, Yubo and {Lai}, Dong},
        title = "{Dynamical tides in eccentric binaries containing massive main-sequence stars: analytical expressions}",
      journal = {\mnras},
         year = 2022,
        month = mar,
       volume = {510},
       number = {4},
        pages = {4943-4951},
          doi = {10.1093/mnras/stab3698},
archivePrefix = {arXiv},
       eprint = {2110.12030},
 primaryClass = {astro-ph.SR},
       adsurl = {https://ui.adsabs.harvard.edu/abs/2022MNRAS.510.4943S}
}

@ARTICLE{Fuller2012,
       author = {{Fuller}, Jim and {Lai}, Dong},
        title = "{Dynamical tides in compact white dwarf binaries: tidal synchronization and dissipation}",
      journal = {\mnras},
         year = 2012,
        month = mar,
       volume = {421},
       number = {1},
        pages = {426-445},
          doi = {10.1111/j.1365-2966.2011.20320.x},
archivePrefix = {arXiv},
       eprint = {1108.4910},
 primaryClass = {astro-ph.SR},
       adsurl = {https://ui.adsabs.harvard.edu/abs/2012MNRAS.421..426F}
}

@article{Storch_2013,
   title={Viscoelastic tidal dissipation in giant planets and formation of hot Jupiters through high-eccentricity migration},
   volume={438},
   ISSN={1365-2966},
   url={http://dx.doi.org/10.1093/mnras/stt2292},
   DOI={10.1093/mnras/stt2292},
   number={2},
   journal={Monthly Notices of the Royal Astronomical Society},
   publisher={Oxford University Press (OUP)},
   author={Storch, Natalia I. and Lai, Dong},
   year={2013},
   month=dec, pages={1526–1534} }

@BOOK{Murray1999,
       author = {{Murray}, Carl D. and {Dermott}, Stanley F.},
        title = "{Solar System Dynamics}",
         year = 1999,
          doi = {10.1017/CBO9781139174817},
       adsurl = {https://ui.adsabs.harvard.edu/abs/1999ssd..book.....M}
}

@ARTICLE{ElBadry23_BH2,
       author = {{El-Badry}, Kareem and {Rix}, Hans-Walter and {Cendes}, Yvette and {Rodriguez}, Antonio C. and {Conroy}, Charlie and {Quataert}, Eliot and {Hawkins}, Keith and {Zari}, Eleonora and {Hobson}, Melissa and {Breivik}, Katelyn and {Rau}, Arne and {Berger}, Edo and {Shahaf}, Sahar and {Seeburger}, Rhys and {Burdge}, Kevin B. and {Latham}, David W. and {Buchhave}, Lars A. and {Bieryla}, Allyson and {Bashi}, Dolev and {Mazeh}, Tsevi and {Faigler}, Simchon},
        title = "{A red giant orbiting a black hole}",
      journal = {\mnras},
         year = 2023,
        month = may,
       volume = {521},
       number = {3},
        pages = {4323-4348},
          doi = {10.1093/mnras/stad799},
archivePrefix = {arXiv},
       eprint = {2302.07880},
 primaryClass = {astro-ph.SR},
       adsurl = {https://ui.adsabs.harvard.edu/abs/2023MNRAS.521.4323E}
}

@ARTICLE{Fuller2012tidalnovae,
       author = {{Fuller}, Jim and {Lai}, Dong},
        title = "{Tidal Novae in Compact Binary White Dwarfs}",
      journal = {\apjl},
         year = 2012,
        month = sep,
       volume = {756},
       number = {1},
          eid = {L17},
        pages = {L17},
          doi = {10.1088/2041-8205/756/1/L17},
archivePrefix = {arXiv},
       eprint = {1206.0470},
 primaryClass = {astro-ph.SR},
       adsurl = {https://ui.adsabs.harvard.edu/abs/2012ApJ...756L..17F}
}

@ARTICLE{Fuller2013MNRAS,
       author = {{Fuller}, Jim and {Lai}, Dong},
        title = "{Dynamical tides in compact white dwarf binaries: helium core white dwarfs, tidal heating and observational signatures}",
      journal = {\mnras},
         year = 2013,
        month = mar,
       volume = {430},
       number = {1},
        pages = {274-287},
          doi = {10.1093/mnras/sts606},
archivePrefix = {arXiv},
       eprint = {1211.0624},
 primaryClass = {astro-ph.SR},
       adsurl = {https://ui.adsabs.harvard.edu/abs/2013MNRAS.430..274F}
}

@article{van_Haaften_2012,
   title={The evolution of ultracompact X-ray binaries},
   volume={537},
   ISSN={1432-0746},
   url={http://dx.doi.org/10.1051/0004-6361/201117880},
   DOI={10.1051/0004-6361/201117880},
   journal={Astronomy and amp; Astrophysics},
   publisher={EDP Sciences},
   author={van Haaften, L. M. and Nelemans, G. and Voss, R. and Wood, M. A. and Kuijpers, J.},
   year={2012},
   month=jan, pages={A104} }

@article{Deloye_2003,
   title={White Dwarf Donors in Ultracompact Binaries: The Stellar Structure of Finite‐Entropy Objects},
   volume={598},
   ISSN={1538-4357},
   url={http://dx.doi.org/10.1086/379063},
   DOI={10.1086/379063},
   number={2},
   journal={The Astrophysical Journal},
   publisher={American Astronomical Society},
   author={Deloye, Christopher J. and Bildsten, Lars},
   year={2003},
   month=dec, pages={1217–1228} }

@article{Sengar_2017,
   title={Novel modelling of ultracompact X-ray binary evolution – stable mass transfer from white dwarfs to neutron stars},
   volume={470},
   ISSN={1745-3933},
   url={http://dx.doi.org/10.1093/mnrasl/slx064},
   DOI={10.1093/mnrasl/slx064},
   number={1},
   journal={Monthly Notices of the Royal Astronomical Society: Letters},
   publisher={Oxford University Press (OUP)},
   author={Sengar, Rahul and Tauris, Thomas M. and Langer, Norbert and Istrate, Alina G.},
   year={2017},
   month=apr, pages={L6–L10} }

@article{Kim_2007,
   title={Chandra
                    Multiwavelength Project X‐Ray Point Source Catalog},
   volume={169},
   ISSN={1538-4365},
   url={http://dx.doi.org/10.1086/511634},
   DOI={10.1086/511634},
   number={2},
   journal={The Astrophysical Journal Supplement Series},
   publisher={American Astronomical Society},
   author={Kim, Minsun and Kim, Dong‐Woo and Wilkes, Belinda J. and Green, Paul J. and Kim, Eunhyeuk and Anderson, Craig S. and Barkhouse, Wayne A. and Evans, Nancy R. and Ivezić, Željko and Karovska, Margarita and Kashyap, Vinay L. and Lee, Myung Gyoon and Maksym, Peter and Mossman, Amy E. and Silverman, John D. and Tananbaum, Harvey D.},
   year={2007},
   month=apr, pages={401–429} }

@ARTICLE{shariat2025triple,
       author = {{Shariat}, Cheyanne and {Naoz}, Smadar and {El-Badry}, Kareem and {Rocha}, Kyle Akira and {Kalogera}, Vicky and {Stephan}, Alexander P. and {Burdge}, Kevin B. and {Angelo}, Isabel},
        title = "{Triple Evolution Pathways to Black Hole Low-mass X-Ray Binaries: Insights from V404 Cygni}",
      journal = {\apj},
         year = 2025,
        month = apr,
       volume = {983},
       number = {2},
          eid = {115},
        pages = {115},
          doi = {10.3847/1538-4357/adbf01},
archivePrefix = {arXiv},
       eprint = {2411.15644},
 primaryClass = {astro-ph.SR},
       adsurl = {https://ui.adsabs.harvard.edu/abs/2025ApJ...983..115S}
}

@ARTICLE{Juri2008MW,
       author = {{Juri{\'c}}, Mario and {Ivezi{\'c}}, {\v{Z}}eljko and {Brooks}, Alyson and {Lupton}, Robert H. and {Schlegel}, David and {Finkbeiner}, Douglas and {Padmanabhan}, Nikhil and {Bond}, Nicholas and {Sesar}, Branimir and {Rockosi}, Constance M. and {Knapp}, Gillian R. and {Gunn}, James E. and {Sumi}, Takahiro and {Schneider}, Donald P. and {Barentine}, J.~C. and {Brewington}, Howard J. and {Brinkmann}, J. and {Fukugita}, Masataka and {Harvanek}, Michael and {Kleinman}, S.~J. and {Krzesinski}, Jurek and {Long}, Dan and {Neilsen}, Jr., Eric H. and {Nitta}, Atsuko and {Snedden}, Stephanie A. and {York}, Donald G.},
        title = "{The Milky Way Tomography with SDSS. I. Stellar Number Density Distribution}",
      journal = {\apj},
         year = 2008,
        month = feb,
       volume = {673},
       number = {2},
        pages = {864-914},
          doi = {10.1086/523619},
archivePrefix = {arXiv},
       eprint = {astro-ph/0510520},
 primaryClass = {astro-ph},
       adsurl = {https://ui.adsabs.harvard.edu/abs/2008ApJ...673..864J}
}

@misc{zhang2024chandrasearchperiodicxray,
      title={A Chandra Search for Periodic X-ray Sources in the Bulge of M31}, 
      author={Jiachang Zhang and Tong Bao and Zhiyuan Li},
      year={2024},
      eprint={2404.07432},
      archivePrefix={arXiv},
      primaryClass={astro-ph.HE},
      url={https://arxiv.org/abs/2404.07432}, 
}

@article{Rahmani_2016,
   title={Star formation laws in the Andromeda galaxy: gas, stars, metals and the surface density of star formation},
   volume={456},
   ISSN={1365-2966},
   url={http://dx.doi.org/10.1093/mnras/stv2951},
   DOI={10.1093/mnras/stv2951},
   number={4},
   journal={Monthly Notices of the Royal Astronomical Society},
   publisher={Oxford University Press (OUP)},
   author={Rahmani, S. and Lianou, S. and Barmby, P.},
   year={2016},
   month=jan, pages={4128–4144} }

@article{Toonen_2018nswd,
   title={The demographics of neutron star – white dwarf mergers: Rates, delay-time distributions, and progenitors},
   volume={619},
   ISSN={1432-0746},
   url={http://dx.doi.org/10.1051/0004-6361/201833164},
   DOI={10.1051/0004-6361/201833164},
   journal={Astronomy and amp; Astrophysics},
   publisher={EDP Sciences},
   author={Toonen, S. and Perets, H. B. and Igoshev, A. P. and Michaely, E. and Zenati, Y.},
   year={2018},
   month=nov, pages={A53} }

@ARTICLE{Chakrabarti2023gaiaBH1,
       author = {{Chakrabarti}, Sukanya and {Simon}, Joshua D. and {Craig}, Peter A. and {Reggiani}, Henrique and {Brandt}, Timothy D. and {Guhathakurta}, Puragra and {Dalba}, Paul A. and {Kirby}, Evan N. and {Chang}, Philip and {Hey}, Daniel R. and {Savino}, Alessandro and {Geha}, Marla and {Thompson}, Ian B.},
        title = "{A Noninteracting Galactic Black Hole Candidate in a Binary System with a Main-sequence Star}",
      journal = {\aj},
         year = 2023,
        month = jul,
       volume = {166},
       number = {1},
          eid = {6},
        pages = {6},
          doi = {10.3847/1538-3881/accf21},
archivePrefix = {arXiv},
       eprint = {2210.05003},
 primaryClass = {astro-ph.GA},
       adsurl = {https://ui.adsabs.harvard.edu/abs/2023AJ....166....6C}
}

@article{Ivanova_2010,
   title={FORMATION OF BLACK HOLE X-RAY BINARIES IN GLOBULAR CLUSTERS},
   volume={717},
   ISSN={1538-4357},
   url={http://dx.doi.org/10.1088/0004-637X/717/2/948},
   DOI={10.1088/0004-637x/717/2/948},
   number={2},
   journal={The Astrophysical Journal},
   publisher={American Astronomical Society},
   author={Ivanova, N. and Chaichenets, S. and Fregeau, J. and Heinke, C. O. and Lombardi, J. C. and Woods, T. E.},
   year={2010},
   month=jun, pages={948–957} }

@article{Ivanova_2017,
   title={Formation of Black Hole X-Ray Binaries with Non-degenerate Donors in Globular Clusters},
   volume={843},
   ISSN={2041-8213},
   url={http://dx.doi.org/10.3847/2041-8213/aa7b76},
   DOI={10.3847/2041-8213/aa7b76},
   number={2},
   journal={The Astrophysical Journal Letters},
   publisher={American Astronomical Society},
   author={Ivanova, Natalia and Rocha, Cassio A. da and Van, Kenny X. and Nandez, Jose L. A.},
   year={2017},
   month=jul, pages={L30} }

@misc{nagarajan2025kick,
      title={Mixed origins: strong natal kicks for some black holes and none for others}, 
      author={Pranav Nagarajan and Kareem El-Badry},
      year={2025},
      eprint={2411.16847},
      archivePrefix={arXiv},
      primaryClass={astro-ph.GA},
      url={https://arxiv.org/abs/2411.16847}, 
}

@article{Fragione_2020,
   title={Electromagnetic transients and gravitational waves from white dwarf disruptions by stellar black holes in triple systems},
   volume={495},
   ISSN={1365-2966},
   url={http://dx.doi.org/10.1093/mnras/staa1192},
   DOI={10.1093/mnras/staa1192},
   number={1},
   journal={Monthly Notices of the Royal Astronomical Society},
   publisher={Oxford University Press (OUP)},
   author={Fragione, Giacomo and Metzger, Brian D and Perna, Rosalba and Leigh, Nathan W C and Kocsis, Bence},
   year={2020},
   month=may, pages={1061–1072} }

@article{Chen_2020UCXB,
   title={Detectability of Ultra-compact X-Ray Binaries as LISA Sources},
   volume={900},
   ISSN={2041-8213},
   url={http://dx.doi.org/10.3847/2041-8213/abae66},
   DOI={10.3847/2041-8213/abae66},
   number={1},
   journal={The Astrophysical Journal Letters},
   publisher={American Astronomical Society},
   author={Chen, Wen-Cong and Liu, Dong-Dong and Wang, Bo},
   year={2020},
   month=aug, pages={L8} }

@article{Suvorov_2021,
   title={Ultra-compact X-ray binaries as dual-line gravitational-wave sources},
   volume={503},
   ISSN={1365-2966},
   url={http://dx.doi.org/10.1093/mnras/stab825},
   DOI={10.1093/mnras/stab825},
   number={4},
   journal={Monthly Notices of the Royal Astronomical Society},
   publisher={Oxford University Press (OUP)},
   author={Suvorov, A G},
   year={2021},
   month=mar, pages={5495–5503} }

@misc{chen2025newpotentialultracompactxray,
      title={New Potential Ultra-compact X-ray Binaries for Space-based Gravitational Wave Detectors From Low-Mass Main-Sequence Companion Channel}, 
      author={Minghua Chen and Jinzhong liu},
      year={2025},
      eprint={2502.11576},
      archivePrefix={arXiv},
      primaryClass={astro-ph.HE},
      url={https://arxiv.org/abs/2502.11576}, 
}

@article{Qin_2023,
   title={Black Hole Ultracompact X-Ray Binaries: Galactic Low-frequency Gravitational Wave Sources},
   volume={944},
   ISSN={1538-4357},
   url={http://dx.doi.org/10.3847/1538-4357/acb340},
   DOI={10.3847/1538-4357/acb340},
   number={1},
   journal={The Astrophysical Journal},
   publisher={American Astronomical Society},
   author={Qin, Ke and Jiang, Long and Chen, Wen-Cong},
   year={2023},
   month=feb, pages={83} }

@article{Heinke_2013,
   title={GALACTIC ULTRACOMPACT X-RAY BINARIES: DISK STABILITY AND EVOLUTION},
   volume={768},
   ISSN={1538-4357},
   url={http://dx.doi.org/10.1088/0004-637X/768/2/184},
   DOI={10.1088/0004-637x/768/2/184},
   number={2},
   journal={The Astrophysical Journal},
   publisher={American Astronomical Society},
   author={Heinke, C. O. and Ivanova, N. and Engel, M. C. and Pavlovskii, K. and Sivakoff, G. R. and Cartwright, T. F. and Gladstone, J. C.},
   year={2013},
   month=apr, pages={184} }

@ARTICLE{Tudor2018,
       author = {{Tudor}, V. and {Miller-Jones}, J.~C.~A. and {Knigge}, C. and {Maccarone}, T.~J. and {Tauris}, T.~M. and {Bahramian}, A. and {Chomiuk}, L. and {Heinke}, C.~O. and {Sivakoff}, G.~R. and {Strader}, J. and {Plotkin}, R.~M. and {Soria}, R. and {Albrow}, M.~D. and {Anderson}, G.~E. and {van den Berg}, M. and {Bernardini}, F. and {Bogdanov}, S. and {Britt}, C.~T. and {Russell}, D.~M. and {Zurek}, D.~R.},
        title = "{HST spectrum and timing of the ultracompact X-ray binary candidate 47 Tuc X9}",
      journal = {\mnras},
         year = 2018,
        month = may,
       volume = {476},
       number = {2},
        pages = {1889-1908},
          doi = {10.1093/mnras/sty284},
archivePrefix = {arXiv},
       eprint = {1802.00161},
 primaryClass = {astro-ph.HE},
       adsurl = {https://ui.adsabs.harvard.edu/abs/2018MNRAS.476.1889T}
}

@article{Ye_2023,
   title={On the Tidal Capture of White Dwarfs by Intermediate-mass Black Holes in Dense Stellar Environments},
   volume={953},
   ISSN={1538-4357},
   url={http://dx.doi.org/10.3847/1538-4357/ace1eb},
   DOI={10.3847/1538-4357/ace1eb},
   number={2},
   journal={The Astrophysical Journal},
   publisher={American Astronomical Society},
   author={Ye, Claire S. and Fragione, Giacomo and Perna, Rosalba},
   year={2023},
   month=aug, pages={141} }

@ARTICLE{Fragos09,
       author = {{Fragos}, T. and {Willems}, B. and {Kalogera}, V. and {Ivanova}, N. and {Rockefeller}, G. and {Fryer}, C.~L. and {Young}, P.~A.},
        title = "{Understanding Compact Object Formation and Natal Kicks. II. The Case of XTE J1118 + 480}",
      journal = {\apj},
         year = 2009,
        month = jun,
       volume = {697},
       number = {2},
        pages = {1057-1070},
          doi = {10.1088/0004-637X/697/2/1057},
archivePrefix = {arXiv},
       eprint = {0809.1588},
 primaryClass = {astro-ph},
       adsurl = {https://ui.adsabs.harvard.edu/abs/2009ApJ...697.1057F}
}

@ARTICLE{Reid14,
       author = {{Reid}, M.~J. and {McClintock}, J.~E. and {Steiner}, J.~F. and {Steeghs}, D. and {Remillard}, R.~A. and {Dhawan}, V. and {Narayan}, R.},
        title = "{A Parallax Distance to the Microquasar GRS 1915+105 and a Revised Estimate of its Black Hole Mass}",
      journal = {\apj},
         year = 2014,
        month = nov,
       volume = {796},
       number = {1},
          eid = {2},
        pages = {2},
          doi = {10.1088/0004-637X/796/1/2},
archivePrefix = {arXiv},
       eprint = {1409.2453},
 primaryClass = {astro-ph.GA},
       adsurl = {https://ui.adsabs.harvard.edu/abs/2014ApJ...796....2R}
}

@INPROCEEDINGS{Mirabel17,
       author = {{Mirabel}, I.~F.},
        title = "{Black holes formed by direct collapse: observational evidences}",
    booktitle = {New Frontiers in Black Hole Astrophysics},
         year = 2017,
       editor = {{Gomboc}, Andreja},
       series = {IAU Symposium},
       volume = {324},
        month = jan,
        pages = {303-306},
          doi = {10.1017/S1743921316012904},
archivePrefix = {arXiv},
       eprint = {1611.09266},
 primaryClass = {astro-ph.HE},
       adsurl = {https://ui.adsabs.harvard.edu/abs/2017IAUS..324..303M}
}

@ARTICLE{Shenar22,
       author = {{Shenar}, Tomer and {Sana}, Hugues and {Mahy}, Laurent and {El-Badry}, Kareem and {Marchant}, Pablo and {Langer}, Norbert and {Hawcroft}, Calum and {Fabry}, Matthias and {Sen}, Koushik and {Almeida}, Leonardo A. and {Abdul-Masih}, Michael and {Bodensteiner}, Julia and {Crowther}, Paul A. and {Gieles}, Mark and {Gromadzki}, Mariusz and {H{\'e}nault-Brunet}, Vincent and {Herrero}, Artemio and {de Koter}, Alex and {Iwanek}, Patryk and {Koz{\l}owski}, Szymon and {Lennon}, Daniel J. and {Ma{\'\i}z Apell{\'a}niz}, Jes{\'u}s and {Mr{\'o}z}, Przemys{\l}aw and {Moffat}, Anthony F.~J. and {Picco}, Annachiara and {Pietrukowicz}, Pawe{\l} and {Poleski}, Rados{\l}aw and {Rybicki}, Krzysztof and {Schneider}, Fabian R.~N. and {Skowron}, Dorota M. and {Skowron}, Jan and {Soszy{\'n}ski}, Igor and {Szyma{\'n}ski}, Micha{\l} K. and {Toonen}, Silvia and {Udalski}, Andrzej and {Ulaczyk}, Krzysztof and {Vink}, Jorick S. and {Wrona}, Marcin},
        title = "{An X-ray-quiet black hole born with a negligible kick in a massive binary within the Large Magellanic Cloud}",
      journal = {Nature Astronomy},
         year = 2022,
        month = jul,
       volume = {6},
        pages = {1085-1092},
          doi = {10.1038/s41550-022-01730-y},
archivePrefix = {arXiv},
       eprint = {2207.07675},
 primaryClass = {astro-ph.HE},
       adsurl = {https://ui.adsabs.harvard.edu/abs/2022NatAs...6.1085S}
}

@ARTICLE{Andrews22,
       author = {{Andrews}, Jeff J. and {Kalogera}, Vicky},
        title = "{Constraining Black Hole Natal Kicks with Astrometric Microlensing}",
      journal = {\apj},
         year = 2022,
        month = may,
       volume = {930},
       number = {2},
          eid = {159},
        pages = {159},
          doi = {10.3847/1538-4357/ac66d6},
archivePrefix = {arXiv},
       eprint = {2203.15156},
 primaryClass = {astro-ph.HE},
       adsurl = {https://ui.adsabs.harvard.edu/abs/2022ApJ...930..159A}
}

@ARTICLE{Kimball23,
       author = {{Kimball}, Chase and {Imperato}, Sam and {Kalogera}, Vicky and {Rocha}, Kyle A. and {Doctor}, Zoheyr and {Andrews}, Jeff J. and {Dotter}, Aaron and {Zapartas}, Emmanouil and {Bavera}, Simone S. and {Kovlakas}, Konstantinos and {Fragos}, Tassos and {Srivastava}, Philipp M. and {Misra}, Devina and {Sun}, Meng and {Xing}, Zepei},
        title = "{A Black Hole Kicked at Birth: MAXI J1305-704}",
      journal = {\apjl},
         year = 2023,
        month = aug,
       volume = {952},
       number = {2},
          eid = {L34},
        pages = {L34},
          doi = {10.3847/2041-8213/ace526},
archivePrefix = {arXiv},
       eprint = {2211.02158},
 primaryClass = {astro-ph.HE},
       adsurl = {https://ui.adsabs.harvard.edu/abs/2023ApJ...952L..34K}
}

@ARTICLE{Dashwood24,
       author = {{Dashwood Brown}, Cordelia and {Gandhi}, Poshak and {Zhao}, Yue},
        title = "{On the natal kick of the black hole X-ray binary H 1705-250}",
      journal = {\mnras},
         year = 2024,
        month = jan,
       volume = {527},
       number = {1},
        pages = {L82-L87},
          doi = {10.1093/mnrasl/slad151},
archivePrefix = {arXiv},
       eprint = {2310.11492},
 primaryClass = {astro-ph.HE},
       adsurl = {https://ui.adsabs.harvard.edu/abs/2024MNRAS.527L..82D}
}

@ARTICLE{Sukhbold16,
       author = {{Sukhbold}, Tuguldur and {Ertl}, T. and {Woosley}, S.~E. and {Brown}, Justin M. and {Janka}, H. -T.},
        title = "{Core-collapse Supernovae from 9 to 120 Solar Masses Based on Neutrino-powered Explosions}",
      journal = {\apj},
         year = 2016,
        month = apr,
       volume = {821},
       number = {1},
          eid = {38},
        pages = {38},
          doi = {10.3847/0004-637X/821/1/38},
archivePrefix = {arXiv},
       eprint = {1510.04643},
 primaryClass = {astro-ph.HE},
       adsurl = {https://ui.adsabs.harvard.edu/abs/2016ApJ...821...38S}
}

@ARTICLE{Woosley95,
       author = {{Woosley}, S.~E. and {Weaver}, Thomas A.},
        title = "{The Evolution and Explosion of Massive Stars. II. Explosive Hydrodynamics and Nucleosynthesis}",
      journal = {\apjs},
         year = 1995,
        month = nov,
       volume = {101},
        pages = {181},
          doi = {10.1086/192237},
       adsurl = {https://ui.adsabs.harvard.edu/abs/1995ApJS..101..181W}
}

@ARTICLE{Burdge24,
       author = {{Burdge}, Kevin B. and {El-Badry}, Kareem and {Kara}, Erin and {Canizares}, Claude and {Chakrabarty}, Deepto and {Frebel}, Anna and {Millholland}, Sarah C. and {Rappaport}, Saul and {Simcoe}, Rob and {Vanderburg}, Andrew},
        title = "{The black hole low-mass X-ray binary V404 Cygni is part of a wide triple}",
      journal = {\nat},
         year = 2024,
        month = nov,
       volume = {635},
       number = {8038},
        pages = {316-320},
          doi = {10.1038/s41586-024-08120-6},
archivePrefix = {arXiv},
       eprint = {2404.03719},
 primaryClass = {astro-ph.HE},
       adsurl = {https://ui.adsabs.harvard.edu/abs/2024Natur.635..316B}
}

@ARTICLE{Hansen97,
       author = {{Hansen}, Brad M.~S. and {Phinney}, E. Sterl},
        title = "{The pulsar kick velocity distribution}",
      journal = {\mnras},
         year = 1997,
        month = nov,
       volume = {291},
       number = {3},
        pages = {569-577},
          doi = {10.1093/mnras/291.3.569},
archivePrefix = {arXiv},
       eprint = {astro-ph/9708071},
 primaryClass = {astro-ph},
       adsurl = {https://ui.adsabs.harvard.edu/abs/1997MNRAS.291..569H}
}

@ARTICLE{Arzoumanian02,
       author = {{Arzoumanian}, Z. and {Chernoff}, D.~F. and {Cordes}, J.~M.},
        title = "{The Velocity Distribution of Isolated Radio Pulsars}",
      journal = {\apj},
         year = 2002,
        month = mar,
       volume = {568},
       number = {1},
        pages = {289-301},
          doi = {10.1086/338805},
archivePrefix = {arXiv},
       eprint = {astro-ph/0106159},
 primaryClass = {astro-ph},
       adsurl = {https://ui.adsabs.harvard.edu/abs/2002ApJ...568..289A}
}

@ARTICLE{Hobbs04,
       author = {{Hobbs}, G. and {Faulkner}, A. and {Stairs}, I.~H. and {Camilo}, F. and {Manchester}, R.~N. and {Lyne}, A.~G. and {Kramer}, M. and {D'Amico}, N. and {Kaspi}, V.~M. and {Possenti}, A. and {McLaughlin}, M.~A. and {Lorimer}, D.~R. and {Burgay}, M. and {Joshi}, B.~C. and {Crawford}, F.},
        title = "{The Parkes multibeam pulsar survey - IV. Discovery of 180 pulsars and parameters for 281 previously known pulsars}",
      journal = {\mnras},
         year = 2004,
        month = aug,
       volume = {352},
       number = {4},
        pages = {1439-1472},
          doi = {10.1111/j.1365-2966.2004.08042.x},
archivePrefix = {arXiv},
       eprint = {astro-ph/0405364},
 primaryClass = {astro-ph},
       adsurl = {https://ui.adsabs.harvard.edu/abs/2004MNRAS.352.1439H}
}

@ARTICLE{Shariat25_10ktriples,
       author = {{Shariat}, Cheyanne and {El-Badry}, Kareem and {Naoz}, Smadar},
        title = "{10,000 Resolved Triples from Gaia: Empirical Constraints on Triple Star Populations}",
      journal = {\pasp},
         year = 2025,
        month = sep,
       volume = {137},
       number = {9},
          eid = {094201},
        pages = {094201},
          doi = {10.1088/1538-3873/adfb30},
archivePrefix = {arXiv},
       eprint = {2506.16513},
 primaryClass = {astro-ph.SR},
       adsurl = {https://ui.adsabs.harvard.edu/abs/2025PASP..137i4201S}
}

@ARTICLE{Tokovinin22,
       author = {{Tokovinin}, Andrei},
        title = "{Resolved Gaia Triples}",
      journal = {\apj},
         year = 2022,
        month = feb,
       volume = {926},
       number = {1},
          eid = {1},
        pages = {1},
          doi = {10.3847/1538-4357/ac4584},
archivePrefix = {arXiv},
       eprint = {2112.11943},
 primaryClass = {astro-ph.SR},
       adsurl = {https://ui.adsabs.harvard.edu/abs/2022ApJ...926....1T}
}

@ARTICLE{Grishin17,
       author = {{Grishin}, Evgeni and {Perets}, Hagai B. and {Zenati}, Yossef and {Michaely}, Erez},
        title = "{Generalized Hill-stability criteria for hierarchical three-body systems at arbitrary inclinations}",
      journal = {\mnras},
         year = 2017,
        month = apr,
       volume = {466},
       number = {1},
        pages = {276-285},
          doi = {10.1093/mnras/stw3096},
archivePrefix = {arXiv},
       eprint = {1609.05912},
 primaryClass = {astro-ph.EP},
       adsurl = {https://ui.adsabs.harvard.edu/abs/2017MNRAS.466..276G}
}

@ARTICLE{Mushkin20,
       author = {{Mushkin}, Jonathan and {Katz}, Boaz},
        title = "{A simple random walk model explains the disruption process of hierarchical, Eccentric three-body systems}",
      journal = {\mnras},
         year = 2020,
        month = oct,
       volume = {498},
       number = {1},
        pages = {665-673},
          doi = {10.1093/mnras/staa2492},
archivePrefix = {arXiv},
       eprint = {2005.03669},
 primaryClass = {astro-ph.SR},
       adsurl = {https://ui.adsabs.harvard.edu/abs/2020MNRAS.498..665M}
}

@ARTICLE{Zhang23,
       author = {{Zhang}, Eric and {Naoz}, Smadar and {Will}, Clifford M.},
        title = "{A Stability Timescale for Non-Hierarchical Three-Body Systems}",
      journal = {arXiv e-prints},
         year = 2023,
        month = jan,
          eid = {arXiv:2301.08271},
        pages = {arXiv:2301.08271},
          doi = {10.48550/arXiv.2301.08271},
archivePrefix = {arXiv},
       eprint = {2301.08271},
 primaryClass = {astro-ph.GA},
       adsurl = {https://ui.adsabs.harvard.edu/abs/2023arXiv230108271Z}
}

@ARTICLE{Toonen2022,
       author = {{Toonen}, S. and {Boekholt}, T.~C.~N. and {Portegies Zwart}, S.},
        title = "{Stellar triples on the edge. Comprehensive overview of the evolution of destabilised triples leading to stellar and binary exotica}",
      journal = {\aap},
         year = 2022,
        month = may,
       volume = {661},
          eid = {A61},
        pages = {A61},
          doi = {10.1051/0004-6361/202141991},
archivePrefix = {arXiv},
       eprint = {2108.04272},
 primaryClass = {astro-ph.SR},
       adsurl = {https://ui.adsabs.harvard.edu/abs/2022A&A...661A..61T}
}

@ARTICLE{Bhaskar21,
       author = {{Bhaskar}, Hareesh and {Li}, Gongjie and {Hadden}, Sam and {Payne}, Matthew J. and {Holman}, Matthew J.},
        title = "{Mildly Hierarchical Triple Dynamics and Applications to the Outer Solar System}",
      journal = {\aj},
         year = 2021,
        month = jan,
       volume = {161},
       number = {1},
          eid = {48},
        pages = {48},
          doi = {10.3847/1538-3881/abcbfc},
archivePrefix = {arXiv},
       eprint = {2008.04335},
 primaryClass = {astro-ph.EP},
       adsurl = {https://ui.adsabs.harvard.edu/abs/2021AJ....161...48B}
}

@ARTICLE{Fragos23,
       author = {{Fragos}, Tassos and {Andrews}, Jeff J. and {Bavera}, Simone S. and {Berry}, Christopher P.~L. and {Coughlin}, Scott and {Dotter}, Aaron and {Giri}, Prabin and {Kalogera}, Vicky and {Katsaggelos}, Aggelos and {Kovlakas}, Konstantinos and {Lalvani}, Shamal and {Misra}, Devina and {Srivastava}, Philipp M. and {Qin}, Ying and {Rocha}, Kyle A. and {Rom{\'a}n-Garza}, Jaime and {Serra}, Juan Gabriel and {Stahle}, Petter and {Sun}, Meng and {Teng}, Xu and {Trajcevski}, Goce and {Tran}, Nam Hai and {Xing}, Zepei and {Zapartas}, Emmanouil and {Zevin}, Michael},
        title = "{POSYDON: A General-purpose Population Synthesis Code with Detailed Binary-evolution Simulations}",
      journal = {\apjs},
         year = 2023,
        month = feb,
       volume = {264},
       number = {2},
          eid = {45},
        pages = {45},
          doi = {10.3847/1538-4365/ac90c1},
archivePrefix = {arXiv},
       eprint = {2202.05892},
 primaryClass = {astro-ph.SR},
       adsurl = {https://ui.adsabs.harvard.edu/abs/2023ApJS..264...45F}
}

@ARTICLE{Paxton11,
       author = {{Paxton}, Bill and {Bildsten}, Lars and {Dotter}, Aaron and {Herwig}, Falk and {Lesaffre}, Pierre and {Timmes}, Frank},
        title = "{Modules for Experiments in Stellar Astrophysics (MESA)}",
      journal = {\apjs},
         year = 2011,
        month = jan,
       volume = {192},
       number = {1},
          eid = {3},
        pages = {3},
          doi = {10.1088/0067-0049/192/1/3},
archivePrefix = {arXiv},
       eprint = {1009.1622},
 primaryClass = {astro-ph.SR},
       adsurl = {https://ui.adsabs.harvard.edu/abs/2011ApJS..192....3P}
}

@ARTICLE{Savonije1986,
       author = {{Savonije}, G.~J. and {de Kool}, M. and {van den Heuvel}, E.~P.~J.},
        title = "{The minimum orbital period for ultra-compact binaries with the helium burning secondaries.}",
      journal = {\aap},
         year = 1986,
        month = jan,
       volume = {155},
        pages = {51-57},
       adsurl = {https://ui.adsabs.harvard.edu/abs/1986A&A...155...51S}
}

@article{Armas_Padilla_2023,
   title={UltraCompCAT: A comprehensive catalogue of ultra-compact and short orbital period X-ray binaries},
   volume={677},
   ISSN={1432-0746},
   url={http://dx.doi.org/10.1051/0004-6361/202346797},
   DOI={10.1051/0004-6361/202346797},
   journal={Astronomy and amp; Astrophysics},
   publisher={EDP Sciences},
   author={Armas Padilla, M. and Corral-Santana, J. M. and Borghese, A. and Cúneo, V. A. and Muñoz-Darias, T. and Casares, J. and Torres, M. A. P.},
   year={2023},
   month=sep, pages={A186} }

@misc{yang2025,
      title={Formation of a Possible Black-hole Ultracompact X-ray Binary with the Shortest Orbital Period}, 
      author={Xing-Peng Yang and Kun Xu and Zhi-Fu Gao and Long Jiang and Wen-Cong Chen},
      year={2025},
      eprint={2505.12689},
      archivePrefix={arXiv},
      primaryClass={astro-ph.HE},
      url={https://arxiv.org/abs/2505.12689}, 
}

@article{Lombardi_Jr__2006, 
   title={Stellar Collisions and Ultracompact X‐Ray Binary Formation},
   volume={640},
   ISSN={1538-4357},
   url={http://dx.doi.org/10.1086/499938},
   DOI={10.1086/499938},
   number={1},
   journal={The Astrophysical Journal},
   publisher={American Astronomical Society},
   author={Lombardi, Jr., J. C. and Proulx, Z. F. and Dooley, K. L. and Theriault, E. M. and Ivanova, N. and Rasio, F. A.},
   year={2006},
   month=mar, pages={441–458} }

@article{Rasio_2000,
   title={Formation of Short-Period Binary Pulsars in Globular Clusters},
   volume={532},
   ISSN={0004-637X},
   url={http://dx.doi.org/10.1086/312555},
   DOI={10.1086/312555},
   number={1},
   journal={The Astrophysical Journal},
   publisher={American Astronomical Society},
   author={Rasio, Frederic A. and Pfahl, Eric D. and Rappaport, Saul},
   year={2000},
   month=mar, pages={L47–L50} }

@misc{burdge2024,
      title={The black hole low mass X-ray binary V404 Cygni is part of a wide hierarchical triple, and formed without a kick}, 
      author={Kevin B. Burdge and Kareem El-Badry and Erin Kara and Claude Canizares and Deepto Chakrabarty and Anna Frebel and Sarah C. Millholland and Saul Rappaport and Rob Simcoe and Andrew Vanderburg},
      year={2024},
      eprint={2404.03719},
      archivePrefix={arXiv},
      primaryClass={astro-ph.HE},
      url={https://arxiv.org/abs/2404.03719}, 
}

@ARTICLE{Maccarone07,
       author = {{Maccarone}, Thomas J. and {Kundu}, Arunav and {Zepf}, Stephen E. and {Rhode}, Katherine L.},
        title = "{A black hole in a globular cluster}",
      journal = {\nat},
         year = 2007,
        month = jan,
       volume = {445},
       number = {7124},
        pages = {183-185},
          doi = {10.1038/nature05434},
archivePrefix = {arXiv},
       eprint = {astro-ph/0701310},
 primaryClass = {astro-ph},
       adsurl = {https://ui.adsabs.harvard.edu/abs/2007Natur.445..183M}
}

@ARTICLE{Chen25,
       author = {{Chen}, Minghua and {Liu}, Jinzhong},
        title = "{New Potential Ultracompact X-Ray Binaries for Space-based Gravitational-wave Detectors from a Low-mass Main-sequence Companion Channel}",
      journal = {\apj},
         year = 2025,
        month = mar,
       volume = {981},
       number = {2},
          eid = {175},
        pages = {175},
          doi = {10.3847/1538-4357/adb618},
archivePrefix = {arXiv},
       eprint = {2502.11576},
 primaryClass = {astro-ph.HE},
       adsurl = {https://ui.adsabs.harvard.edu/abs/2025ApJ...981..175C}
}

@ARTICLE{Sana14,
       author = {{Sana}, H. and {Le Bouquin}, J. -B. and {Lacour}, S. and {Berger}, J. -P. and {Duvert}, G. and {Gauchet}, L. and {Norris}, B. and {Olofsson}, J. and {Pickel}, D. and {Zins}, G. and {Absil}, O. and {de Koter}, A. and {Kratter}, K. and {Schnurr}, O. and {Zinnecker}, H.},
        title = "{Southern Massive Stars at High Angular Resolution: Observational Campaign and Companion Detection}",
      journal = {\apjs},
         year = 2014,
        month = nov,
       volume = {215},
       number = {1},
          eid = {15},
        pages = {15},
          doi = {10.1088/0067-0049/215/1/15},
archivePrefix = {arXiv},
       eprint = {1409.6304},
 primaryClass = {astro-ph.SR},
       adsurl = {https://ui.adsabs.harvard.edu/abs/2014ApJS..215...15S}
}

@INPROCEEDINGS{Offner23,
       author = {{Offner}, S.~S.~R. and {Moe}, M. and {Kratter}, K.~M. and {Sadavoy}, S.~I. and {Jensen}, E.~L.~N. and {Tobin}, J.~J.},
        title = "{The Origin and Evolution of Multiple Star Systems}",
    booktitle = {Protostars and Planets VII},
         year = 2023,
       editor = {{Inutsuka}, S. and {Aikawa}, Y. and {Muto}, T. and {Tomida}, K. and {Tamura}, M.},
       series = {Astronomical Society of the Pacific Conference Series},
       volume = {534},
        month = jul,
        pages = {275},
          doi = {10.48550/arXiv.2203.10066},
archivePrefix = {arXiv},
       eprint = {2203.10066},
 primaryClass = {astro-ph.SR},
       adsurl = {https://ui.adsabs.harvard.edu/abs/2023ASPC..534..275O}
}

@ARTICLE{Naoz16_LMXB,
       author = {{Naoz}, Smadar and {Fragos}, Tassos and {Geller}, Aaron and {Stephan}, Alexander P. and {Rasio}, Frederic A.},
        title = "{Formation of Black Hole Low-mass X-Ray Binaries in Hierarchical Triple Systems}",
      journal = {\apjl},
         year = 2016,
        month = may,
       volume = {822},
       number = {2},
          eid = {L24},
        pages = {L24},
          doi = {10.3847/2041-8205/822/2/L24},
archivePrefix = {arXiv},
       eprint = {1510.02093},
 primaryClass = {astro-ph.HE},
       adsurl = {https://ui.adsabs.harvard.edu/abs/2016ApJ...822L..24N}
}

@ARTICLE{Nelson1986,
       author = {{Nelson}, L.~A. and {Rappaport}, S.~A. and {Joss}, P.~C.},
        title = "{The Evolution of Ultrashort Period Binary Systems}",
      journal = {\apj},
         year = 1986,
        month = may,
       volume = {304},
        pages = {231},
          doi = {10.1086/164156},
       adsurl = {https://ui.adsabs.harvard.edu/abs/1986ApJ...304..231N}
}

@article{in_t_Zand_2007,
   title={Six new candidate ultracompact X-ray binaries},
   volume={465},
   ISSN={1432-0746},
   url={http://dx.doi.org/10.1051/0004-6361:20066678},
   DOI={10.1051/0004-6361:20066678},
   number={3},
   journal={Astronomy and amp; Astrophysics},
   publisher={EDP Sciences},
   author={in ’t Zand, J. J. M. and Jonker, P. G. and Markwardt, C. B.},
   year={2007},
   month=jan, pages={953–963} }

@article{Ruan_2020,
   title={The LISA–Taiji network},
   volume={4},
   ISSN={2397-3366},
   url={http://dx.doi.org/10.1038/s41550-019-1008-4},
   DOI={10.1038/s41550-019-1008-4},
   number={2},
   journal={Nature Astronomy},
   publisher={Springer Science and Business Media LLC},
   author={Ruan, Wen-Hong and Liu, Chang and Guo, Zong-Kuan and Wu, Yue-Liang and Cai, Rong-Gen},
   year={2020},
   month=feb, pages={108–109} }

@article{MillerJones2015,
   title={Deep radio imaging of 47 Tuc identifies the peculiar X-ray source X9 as a new black hole candidate},
   volume={453},
   ISSN={1365-2966},
   url={http://dx.doi.org/10.1093/mnras/stv1869},
   DOI={10.1093/mnras/stv1869},
   number={4},
   journal={Monthly Notices of the Royal Astronomical Society},
   publisher={Oxford University Press (OUP)},
   author={Miller-Jones, J. C. A. and Strader, J. and Heinke, C. O. and Maccarone, T. J. and van den Berg, M. and Knigge, C. and Chomiuk, L. and Noyola, E. and Russell, T. D. and Seth, A. C. and Sivakoff, G. R.},
   year={2015},
   month=sep, pages={3919–3932} }

@article{Bahramian2017,
   title={The ultracompact nature of the black hole candidate X-ray binary 47 Tuc X9},
   volume={467},
   ISSN={1365-2966},
   url={http://dx.doi.org/10.1093/mnras/stx166},
   DOI={10.1093/mnras/stx166},
   number={2},
   journal={Monthly Notices of the Royal Astronomical Society},
   publisher={Oxford University Press (OUP)},
   author={Bahramian, Arash and Heinke, Craig O. and Tudor, Vlad and Miller-Jones, James C. A. and Bogdanov, Slavko and Maccarone, Thomas J. and Knigge, Christian and Sivakoff, Gregory R. and Chomiuk, Laura and Strader, Jay and Garcia, Javier A. and Kallman, Timothy},
   year={2017},
   month=feb, pages={2199–2216} }

@article{Church2017,
   title={Formation Constraints Indicate a Black Hole Accretor in 47 Tuc X9},
   volume={851},
   ISSN={2041-8213},
   url={http://dx.doi.org/10.3847/2041-8213/aa9aeb},
   DOI={10.3847/2041-8213/aa9aeb},
   number={1},
   journal={The Astrophysical Journal Letters},
   publisher={American Astronomical Society},
   author={Church, Ross P. and Strader, Jay and Davies, Melvyn B. and Bobrick, Alexey},
   year={2017},
   month=dec, pages={L4} }

@ARTICLE{Podsiadlowski2002,
       author = {{Podsiadlowski}, Ph. and {Rappaport}, S. and {Pfahl}, E.~D.},
        title = "{Evolutionary Sequences for Low- and Intermediate-Mass X-Ray Binaries}",
      journal = {\apj},
         year = 2002,
        month = feb,
       volume = {565},
       number = {2},
        pages = {1107-1133},
          doi = {10.1086/324686},
archivePrefix = {arXiv},
       eprint = {astro-ph/0107261},
 primaryClass = {astro-ph},
       adsurl = {https://ui.adsabs.harvard.edu/abs/2002ApJ...565.1107P}
}

@article{Levesque2005,
   title={The Effective Temperature Scale of Galactic Red Supergiants: Cool, but Not as Cool as We Thought},
   volume={628},
   ISSN={1538-4357},
   url={http://dx.doi.org/10.1086/430901},
   DOI={10.1086/430901},
   number={2},
   journal={The Astrophysical Journal},
   publisher={American Astronomical Society},
   author={Levesque, Emily M. and Massey, Philip and Olsen, K. A. G. and Plez, Bertrand and Josselin, Eric and Maeder, Andre and Meynet, Georges},
   year={2005},
   month=aug, pages={973–985} }

@article{Romagnolo2023,
   title={The role of stellar expansion on the formation of gravitational wave sources},
   volume={525},
   ISSN={1365-2966},
   url={http://dx.doi.org/10.1093/mnras/stad2366},
   DOI={10.1093/mnras/stad2366},
   number={1},
   journal={Monthly Notices of the Royal Astronomical Society},
   publisher={Oxford University Press (OUP)},
   author={Romagnolo, A and Belczynski, K and Klencki, J and Agrawal, P and Shenar, T and Szécsi, D},
   year={2023},
   month=aug, pages={706–720} }

@ARTICLE{Podsiadlowski2010,
       author = {{Podsiadlowski}, Philipp and {Ivanova}, Natasha and {Justham}, Stephen and {Rappaport}, Saul},
        title = "{Explosive common-envelope ejection: implications for gamma-ray bursts and low-mass black-hole binaries}",
      journal = {\mnras},
         year = 2010,
        month = aug,
       volume = {406},
       number = {2},
        pages = {840-847},
          doi = {10.1111/j.1365-2966.2010.16751.x},
archivePrefix = {arXiv},
       eprint = {1004.0249},
 primaryClass = {astro-ph.HE},
       adsurl = {https://ui.adsabs.harvard.edu/abs/2010MNRAS.406..840P}
}

@misc{Ivanova2011,
      title={Common envelope: the progress and the pitfalls}, 
      author={Natalia Ivanova},
      year={2011},
      eprint={1108.1226},
      archivePrefix={arXiv},
      primaryClass={astro-ph.SR},
      url={https://arxiv.org/abs/1108.1226}, 
}

@article{Ivanova2015,
   title={On the role of recombination in common-envelope ejections},
   volume={447},
   ISSN={0035-8711},
   url={http://dx.doi.org/10.1093/mnras/stu2582},
   DOI={10.1093/mnras/stu2582},
   number={3},
   journal={Monthly Notices of the Royal Astronomical Society},
   publisher={Oxford University Press (OUP)},
   author={Ivanova, N. and Justham, S. and Podsiadlowski, Ph.},
   year={2015},
   month=jan, pages={2181–2197} }

@misc{stegmann2025,
      title={Spin-orbit misalignment and residual eccentricity are evidence that neutron star-black hole mergers form through triple star evolution}, 
      author={Jakob Stegmann and Jakub Klencki},
      year={2025},
      eprint={2506.09121},
      archivePrefix={arXiv},
      primaryClass={astro-ph.HE},
      url={https://arxiv.org/abs/2506.09121}, 
}

@misc{xuan2025GC,
      title={Localizing Dynamically-Formed Black Hole Binaries in Milky Way Globular Clusters with LISA}, 
      author={Zeyuan Xuan and Kyle Kremer and Smadar Naoz},
      year={2025},
      eprint={2501.18682},
      archivePrefix={arXiv},
      primaryClass={astro-ph.HE},
      url={https://arxiv.org/abs/2501.18682}, 
}

@misc{lau2025,
      title={Gravitational Radiation-Driven Chaotic Tide in a White Dwarf-Massive Black Hole Binary as a Source of Repeating X-ray Transients}, 
      author={Shu Yan Lau and Hang Yu},
      year={2025},
      eprint={2506.10163},
      archivePrefix={arXiv},
      primaryClass={astro-ph.HE},
      url={https://arxiv.org/abs/2506.10163}, 
}

@inbook{Bahramian_2023,
   title={Low-Mass X-ray Binaries},
   ISBN={9789811645440},
   url={http://dx.doi.org/10.1007/978-981-16-4544-0_94-1},
   DOI={10.1007/978-981-16-4544-0_94-1},
   booktitle={Handbook of X-ray and Gamma-ray Astrophysics},
   publisher={Springer Nature Singapore},
   author={Bahramian, Arash and Degenaar, Nathalie},
   year={2023},
   pages={1–62} }

@misc{pelisoli2025observationaloverviewwhitedwarf,
      title={An observational overview of white dwarf stars}, 
      author={Ingrid Pelisoli and Jamie Williams},
      year={2025},
      eprint={2502.19496},
      archivePrefix={arXiv},
      primaryClass={astro-ph.SR},
      url={https://arxiv.org/abs/2502.19496}, 
}

@ARTICLE{Zhang2024,
       author = {{Zhang}, Jiachang and {Bao}, Tong and {Li}, Zhiyuan},
        title = "{A Chandra search for periodic X-ray sources in the bulge of M31}",
      journal = {\mnras},
         year = 2024,
        month = may,
       volume = {530},
       number = {2},
        pages = {2096-2113},
          doi = {10.1093/mnras/stae1002},
archivePrefix = {arXiv},
       eprint = {2404.07432},
 primaryClass = {astro-ph.HE},
       adsurl = {https://ui.adsabs.harvard.edu/abs/2024MNRAS.530.2096Z}
}

@ARTICLE{Bobrick2017,
       author = {{Bobrick}, Alexey and {Davies}, Melvyn B. and {Church}, Ross P.},
        title = "{Mass transfer in white dwarf-neutron star binaries}",
      journal = {\mnras},
         year = 2017,
        month = may,
       volume = {467},
       number = {3},
        pages = {3556-3575},
          doi = {10.1093/mnras/stx312},
archivePrefix = {arXiv},
       eprint = {1702.02377},
 primaryClass = {astro-ph.HE},
       adsurl = {https://ui.adsabs.harvard.edu/abs/2017MNRAS.467.3556B}
}

@ARTICLE{Tokovinin14,
       author = {{Tokovinin}, Andrei},
        title = "{From Binaries to Multiples. II. Hierarchical Multiplicity of F and G Dwarfs}",
      journal = {\aj},
         year = 2014,
        month = apr,
       volume = {147},
       number = {4},
          eid = {87},
        pages = {87},
          doi = {10.1088/0004-6256/147/4/87},
archivePrefix = {arXiv},
       eprint = {1401.6827},
 primaryClass = {astro-ph.SR},
       adsurl = {https://ui.adsabs.harvard.edu/abs/2014AJ....147...87T}
}

@ARTICLE{Winters19,
       author = {{Winters}, Jennifer G. and {Henry}, Todd J. and {Jao}, Wei-Chun and {Subasavage}, John P. and {Chatelain}, Joseph P. and {Slatten}, Ken and {Riedel}, Adric R. and {Silverstein}, Michele L. and {Payne}, Matthew J.},
        title = "{The Solar Neighborhood. XLV. The Stellar Multiplicity Rate of M Dwarfs Within 25 pc}",
      journal = {\aj},
         year = 2019,
        month = jun,
       volume = {157},
       number = {6},
          eid = {216},
        pages = {216},
          doi = {10.3847/1538-3881/ab05dc},
archivePrefix = {arXiv},
       eprint = {1901.06364},
 primaryClass = {astro-ph.SR},
       adsurl = {https://ui.adsabs.harvard.edu/abs/2019AJ....157..216W}
}

@ARTICLE{Moe21,
       author = {{Moe}, Maxwell and {Kratter}, Kaitlin M.},
        title = "{Impact of binary stars on planet statistics - I. Planet occurrence rates and trends with stellar mass}",
      journal = {\mnras},
         year = 2021,
        month = nov,
       volume = {507},
       number = {3},
        pages = {3593-3611},
          doi = {10.1093/mnras/stab2328},
archivePrefix = {arXiv},
       eprint = {1912.01699},
 primaryClass = {astro-ph.EP},
       adsurl = {https://ui.adsabs.harvard.edu/abs/2021MNRAS.507.3593M}
}

@ARTICLE{Hills1983,
       author = {{Hills}, J.~G.},
        title = "{The effects of sudden mass loss and a random kick velocity produced in a supernova explosion on the dynamics of a binary star of arbitrary orbital eccentricity. Applications to X-ray binaries and to the binarypulsars.}",
      journal = {\apj},
         year = 1983,
        month = apr,
       volume = {267},
        pages = {322-333},
          doi = {10.1086/160871},
       adsurl = {https://ui.adsabs.harvard.edu/abs/1983ApJ...267..322H}
}

@ARTICLE{Shariat2023,
       author = {{Shariat}, Cheyanne and {Naoz}, Smadar and {Hansen}, Bradley M.~S. and {Angelo}, Isabel and {Michaely}, Erez and {Stephan}, Alexander P.},
        title = "{Dynamical Evolution of White Dwarfs in Triples in the Era of Gaia}",
      journal = {\apjl},
         year = 2023,
        month = sep,
       volume = {955},
       number = {1},
          eid = {L14},
        pages = {L14},
          doi = {10.3847/2041-8213/acf76b},
archivePrefix = {arXiv},
       eprint = {2306.13130},
 primaryClass = {astro-ph.SR},
       adsurl = {https://ui.adsabs.harvard.edu/abs/2023ApJ...955L..14S}
}

@ARTICLE{Ivanova2008,
       author = {{Ivanova}, N. and {Heinke}, C.~O. and {Rasio}, F.~A. and {Belczynski}, K. and {Fregeau}, J.~M.},
        title = "{Formation and evolution of compact binaries in globular clusters - II. Binaries with neutron stars}",
      journal = {\mnras},
         year = 2008,
        month = may,
       volume = {386},
       number = {1},
        pages = {553-576},
          doi = {10.1111/j.1365-2966.2008.13064.x},
archivePrefix = {arXiv},
       eprint = {0706.4096},
 primaryClass = {astro-ph},
       adsurl = {https://ui.adsabs.harvard.edu/abs/2008MNRAS.386..553I}
}

@article{Lu_2019,
   title={Supernovae Kicks in hierarchical triple systems},
   ISSN={1365-2966},
   url={http://dx.doi.org/10.1093/mnras/stz036},
   DOI={10.1093/mnras/stz036},
   journal={Monthly Notices of the Royal Astronomical Society},
   publisher={Oxford University Press (OUP)},
   author={Lu, Cicero X and Naoz, Smadar},
   year={2019},
   month=jan }

@article{Kalogera_2000,
   title={Spin‐Orbit Misalignment in Close Binaries with Two Compact Objects},
   volume={541},
   ISSN={1538-4357},
   url={http://dx.doi.org/10.1086/309400},
   DOI={10.1086/309400},
   number={1},
   journal={The Astrophysical Journal},
   publisher={American Astronomical Society},
   author={Kalogera, Vassiliki},
   year={2000},
   month=sep, pages={319–328} }
\end{document}